\documentclass[12pt,preprint]{aastex}

\shorttitle{Mg-$\sigma$}
\shortauthors{Worthey \& Collobert}

\begin{document}

\title{The Mg-$\sigma$ Relation and the Genesis 
	of Early-Type Galaxies}

\author{Guy Worthey \& Maela Collobert\altaffilmark{1}}
\affil{Washington State University}
\affil{1245 Webster Hall, Pullman, WA 99163-2814}
\email{gworthey@wsu.edu}
\altaffiltext{1}{Universit\'e Paris-Sud, France}

\begin{abstract}

Available data on the magnesium - velocity dispersion (Mg-$\sigma$)
relation for $\sim$2000 early-type galaxies is collected and
compared. As noted previously, the Mg residuals from a fitted line are
roughly Gaussian near the mean but have an asymmetric blue tail,
probably from subpopulations of relatively young stars. We define
statistics for scatter and asymmetry of scatter in the Mg dimension
and find impressive uniformity among data sets. We construct models of
galaxy formation built to be as unbiased as possible toward the
question of the importance of mergers in the formation of early type
galaxies. The observational constraints (Mg-$\sigma$ width, asymmetry,
and mean Mg strength, plus mean age and width of abundance
distribution) are severe enough to eliminate almost all of
models. Among the casualties are: models with merger rates
proportional to $(1+z)^{n}$ with $n>0$, models that assume early
formation followed by recent drizzling of new stars, merger-only
models where the number of mergers exceeds $\approx$80, merger-only
models with less than $\approx$20 mergers, and models with a cold dark
matter power spectrum even with biasing included. The most successful
models were those with merger probability constant with time, with the
number of mergers needed to form the galaxy around 40. These models
are characterized by mean light-weighted ages of 7-10 Gyr (consistent
with spectroscopic studies), a narrow abundance distribution, and a
lookback time behavior nearly indistinguishable from passive evolution of old
stellar populations.

\end{abstract}

Galaxies : Abundance --- Galaxies : Elliptical --- Galaxies : Evolution --- 
Galaxies : formation --- Galaxies : Stellar content.
\section{Introduction}

The process of formation of early-type galaxies is still largely
unknown. Under study since \citet{Larson(1975)}, the choice between
formation of elliptical galaxies by a single brief collapse or via
merging is still disputed in the astronomical literature.

Some properties of early-type galaxies are very uniform, suggesting a
homogeneous and probably ancient origin. Elliptical and S0 galaxies
have smooth light profiles and very low gas and dust content, like
overgrown globular clusters. As a class, they display scaling
relations such as a tight Fundamental Plane (FP: \citet{djor87,
dress87}) among the variables luminosity, size, and velocity
dispersion, a narrow Mg-$\sigma$ relation between integrated starlight
absorption line strength and velocity dispersion \citet{bend93,
Zieg97}, and an orderly color-magnitude relation.  Many authors note
that elliptical galaxies nearby and at intermediate redshift have
properties consistent with old, passively evolving systems
\citep{Stand98,vandok96,kel97,ellis97,koda97,koda98,bend98, vandok98}.

Confusingly, there is also a great deal of evidence that favors a
messier, ongoing formation process for early-type galaxies. Some
galaxy mergers are caught in the act \citep{toom77} with the
theoretical expectation that the remnant will soon relax to resemble
an elliptical, with an exponential light profile. The tails, ripples,
shells, and other morphological aftereffects of merging are seen in
many nearby elliptical galaxies \citep{sch1, sch2, goud01, barb02,
mark02}.  Measurements of the integrated stellar absorption features
compared with stellar population models indicate light-weighted mean
ages for the near-nuclear regions of elliptical galaxies that range
from ancient through a median of $\sim$7 Gyr to a few that appear less
than 1 Gyr old \citep{gonza93,wor97,terle02}. Young stellar
populations are much brighter than old populations, so that a
relatively modest burst of young stars may be able to skew the mean
age of an essentially old galaxy to appear much younger than a
mass-weighted mean age, but the presence of any youthful subpopulation
contradicts the uniform-and-old hypothesis.

With this paper we attempt to bring clarity to the issue.  To
simplify, we consider only the relation between the central velocity
dispersion $\sigma$ and the strength of the integrated stellar Mg and
MgH features around 5100\AA. The advantages of this relation are its
distance independence and its small scatter \citep{jorg96,bend98}. One
disadvantage is that age and metallicity both have a similar effect on
the Mg strength \citep{wor94,forb01,coll99}: an older galaxy has a
stronger Mg feature, but Mg strength can also be increased by Mg
abundance. It is clear that the origin of the relation itself is
primarily one of abundance: larger galaxies have more heavy elements.
Intrinsic scatter exits in the $Mg_2$-$\sigma_0$ relation ($\sigma_0$
refers to an aperture-corrected central velocity dispersion). This
scatter is not correlated with the appearance of the galaxies, with
the degree of velocity anisotropy, with their environment
\citep{dress87,burs90}, or with deviation of the objects within (or
perpendiculiar) to the FP. \citet{bend93} intrepreted this as a
combination of age and/or metallicity differencies among ellipticals
with similar luminosity, and \citet{coll99} find that both mean age
and metallicity vary, probably in a correlated manner with younger-appearing
galaxies tending to be somewhat more metal-rich. This correlation is
also seen with Balmer-metal indicators \citep{wor95,terle02} 

The residuals of $Mg_2$ have a Gaussian core, but the galaxies with
weaker $Mg_2$ at given $\sigma_0$ form a tail. That is, the residual
distribution is skewed with an excess of Mg-weak galaxies
\citep{bend93}. This is suggestive of late star formation: A youthful
galaxy will have weak Mg strength from the presence of bright, hot
stars with weak Mg strength and will appear as a low outlier.

We test two opposing hypotheses. In both cases, the Mg-$\sigma$
relation is a mass-abundance correlation. In the ``ancient formation''
hypothesis, elliptical galaxies formed at the epoch of galaxy
formation at high redshift (we assume $z_f$=5). The Mg-$\sigma$
relation was put in place at that time. Furthermore, the symmetric
portion of the scatter was put in place at formation. (We could be
even more extreme and suggest that the asymmetric part of scatter was
also imprinted at formation by an asymmetric abundance distribution,
but the implication of this is trivial: except for the effect of
passive stellar evolution the Mg-$\sigma$ relation would remain
unchanged back to $z_f$.) In the ``ancient formation'' hypothesis,
mergers, as inconsequential as possible, cause occasional bursts of
star formation that drive a few galaxies to weak Mg strength and cause
the asymmetry of the observed relation.
In the ``maximum merger'' hypothesis, the underlying Mg-$\sigma$
relation is very tight, and all scatter, symmetric or not, is caused
by merging events (convolved with observational error). 

To compare these two views, we first collect and analyze published
Mg-$\sigma$ data sets of E and S0 galaxies from field, group, and
cluster environments. We use these data to find robust statistics to
help us evaluate the expected asymmetrical scatter. In $\S$3 we
describe special-purpose models that predict Mg-$\sigma$ relations
given input parameters such as the number of mergers and gas fraction
of each merger. Results are discussed in $\S$4, and a summary of
conclusions is given in $\S$5.

\section{Observational Data}

We analyse the Mg-$\sigma$ relation in 7 different data sets for E and
S0 galaxies.  There are three Lick/IDS system Mg indices: Mg$_1$,
Mg$_2$, Mg$b$ \citep{wor&94}, plus the variant $<$Mg$_2>$ index
introduced by \citet{Davies87}, and defined by :
\begin{displaymath}
<Mg_2> = 0.6\,Mg_2 + 0.4\,Mg^{'}, \ {\rm with}\  Mg^{'} = 0.03 + 2.1\,Mg_1 - 62 Mg^4_1.
\end{displaymath}  
$<$Mg$_2>$ uses the additional signal available in Mg$_1$ to improve
the precision of the Mg$_2$ index. There does not appear to be
any substantial mismatch between $<$Mg$_2>$ and Mg$_2$ definitions in
ultimate index behavior, although differences are detectable
\citep{trag98}. We summarize the literature data that we collected
with a few words and a table of characteristics in table \ref{tab-1}.
 
Data set 1 is a subset of data extracted from \citet{trag98}, it
consists of 256 galaxies observed at the Lick Observatory between 1972
and 1984. Average errors for Mg$_2$, Mgb, and $<$Mg$_2>$ indices were
0.008 mag, 0.23 \AA, 0.006 mag, respectively. For velocity dispersion
the error is less than 10 $\%$. The advantage of these data is that
they define the Lick system and are very homogeneous. The disadvantage
is that no aperture corrections for distance have been included. This
appears to have no measureable consequence (see Table \ref{tab-1}).

Data set 2 is part of the ENEAR database from \citet{bernar98} that
does not overlap any other samples. It is separated
into 3 data sets according to whether the galaxies were assigned to
field, group or cluster environments. Measurements of $\sigma$ are
accurate to 5\%-13\%, Mg$_2$ to 0.005-0.011 mag.

Data set 3: \citet{dress91} have presented 136 cluster and group
elliptical and S0 galaxies in the direction of the large-scale
streaming flow attributed to the great attractor. The typical errors
are 0.05 dex for log{$\sigma$} and 0.017 mag for Mg$_2$.

Data set 4: The 528 galaxies available with the Mg$_2$ index in
\citet{hudson01} are from 56 galaxy clusters.  Uncertainties on
log$\sigma$ are 5\% and 0.009 mag for Mg$_2$.

Data set 5: We culled almost 150 galaxies taken from the EFAR data 
\citep{weg99}. The whole sample is much larger but we
considered only those galaxies that had errors of 7\% or less in
$\sigma$. Errors for Mg$_2$ were 0.015 mag and for Mg$b$ 0.37 \AA.

Data set 6: \citet{Davies87} (the 7 Samurai) computed $\sigma$ and
lines indices for 600 galaxies from all environments.  The scatter of
measurements indicates an uncertainty of 10 \% and 0.009 mag for
$\sigma$ and $<$Mg$_2>$, respectively. We use an updated 7 Samurai
data set kindly provided by D.Burstein (1999, private communication).

Data set 7: This data set of 650 ellipticals and SO galaxies was
extracted from \citet{pru96}. It is a literature compilation with
considerable overlap with the 7 Samurai data set but with different
procedures for homogenization. The data are in $\sigma$ (same error:
10\%), and in Mg$_2$ (error : 0.009 mag).

With the data in hand we then sought suitable statistics to describe
the scatter and asymmetry of the residuals from a linear fit to the
data. The linear fit itself was tried several ways.  A line fit via
least-squares method was too sensitive to the presence of
outliers. Even when we clipped the outliers in a rejection loop we
judged that the resulting line was too unstable.  Finally, we binned
the data in log$\sigma$, found the median Mg index value of each bin,
weighted each datum by $\frac{1}{\sqrt{N}}$, and performed a
least-squares fit on the array of medians. This method is
outlier-insensitive and very stable.  Unless specified otherwise in
this paper, the fit was computed only for $150 < \sigma < 320$ km/s;
medium-sized and large galaxies.  In table~\ref{tab-1}, we report the
number of galaxies really used for our fit.

\begin{figure}
\epsscale{0.7}
\plotone{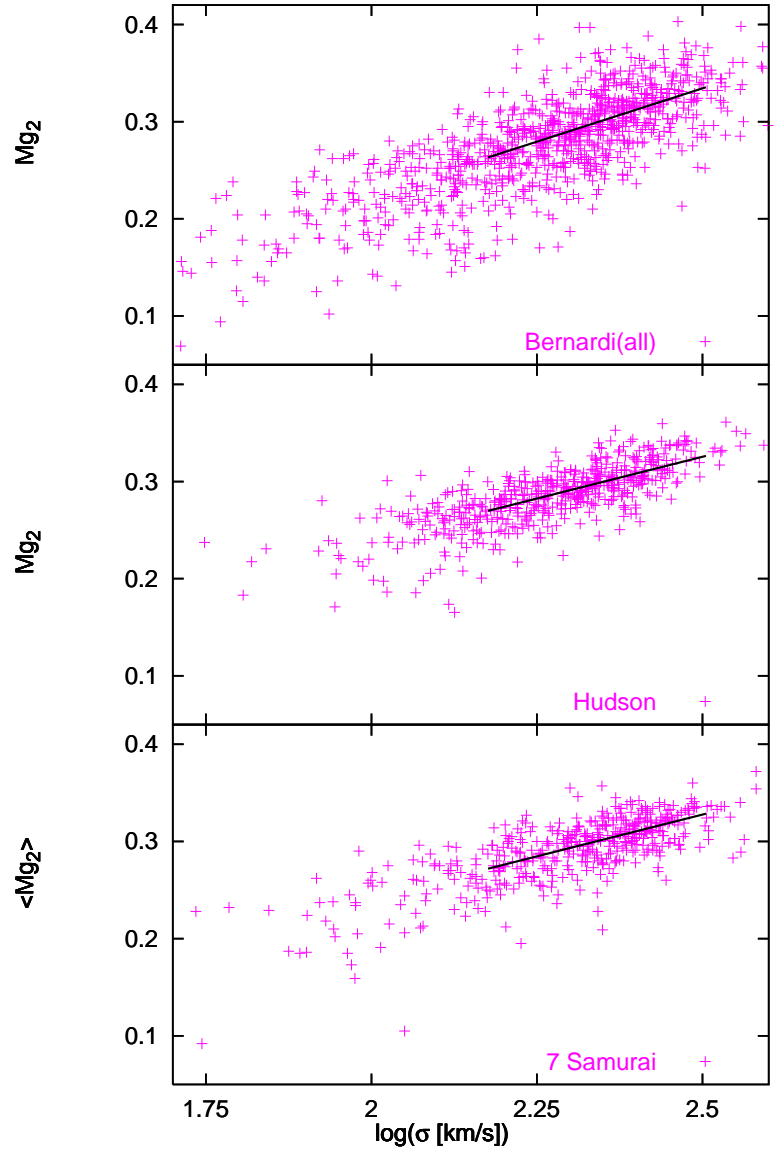}
\figcaption{Observational  Mg$_2$-$\sigma$ relation for 3 of our data sets :
Bernardi (all galaxies), Hudson, and 7 Samurai. The median fit line is 
shown within the limits we chose for calculating statistics : 
150 $<\, \sigma\, <$ 320 km/s. \label{fig1}}
\end{figure}

There is a small amount of systematic drift among the data sets.
The Table~\ref{tab-1} column marked ``Mg 300'' is the value of the Mg
index fit for $\sigma = 300$ km/s. This quantity is
0.327 $\pm$ 0.039 for Mg$_2$, 5.1 $\pm$ 0.085 for Mg$b$, and 0.333
$\pm$ 0.007 for $<$Mg$_2>$, averaged over the data sets.
For illustration, we plot three data sets and the computed median line
in Figure~\ref{fig1}. 

\placefigure{fig1}

Once the line was fit, we sought statistics to characterize the
distribution of Mg index residuals in terms of both width and
asymmetry.  For width, the standard deviation has the advantage that
everyone is familiar with it, but it is quite sensitive to outliers due 
to its $x^2$ dependence.  We prefer the average deviation, or ADev (mean 
absolute deviation). It is an estimator of width less sensitive
to outliers than the standard deviation. It is defined by :
\begin{displaymath}
ADev(x_1\dots x_N) = \frac{1}{N}\,\sum^N _{j=1} \vert x_j-\bar{x} \vert 
\end{displaymath}
but in this case we substitute the median-fit line value $x_{med}$ for 
$\bar{x}$.

For measuring asymmetric tails we need another statistic.  We
considered the skewness, a dimensionless number which describes only
the shape of the distribution. We take the usual definition :
\begin{displaymath}
Skew(x_1\,\dots\,x_N) = \frac{1}{N}\sum ^N _{j=1} \left( \begin{array}{c} 
\frac{x_j-\bar{x}}{\sigma} \end{array} \right) ^{3}
\end{displaymath}
where $\sigma$ is the distribution's standard deviation. A positive value 
of skewness signifies a distribution with an asymmetric tail extending 
out towards more positive $x$. A negative value signifies a 
distribution whose tail extends out towards more negative $x$. We expect a 
negative skewness if galaxies scatter preferentially 
toward weaker Mg index values.

Because we wish to be outlier-insensitive, we also define an ``average
deviation ratio'' (ADR) of the ADevs above and below the fitted line, that 
avoids the $x^3$ dependence of
traditional skewness. Since we expect more weak-lined galaxies than
strong-lined, we define the ADR so that it will be greater than one
for asymmetries toward weak Mg index values.
\begin{displaymath}
ADR_{Mg_2} = \frac{\sum ^N _{j=1} \vert x_j-\bar{x} \vert_{``bottom''}}
{\sum ^N _{j=1} \vert x_j-\bar{x} \vert_{``top''} }
\end{displaymath}

The results are summarised in table~\ref{tab-1}. The value
of the standard deviation is relatively stable at 0.027 $\pm$ 0.003
for Mg$_2$, 0.474 $\pm$ 0.052 for Mg$b$ and 0.035 $\pm$ 0.003 for
$<$Mg$_2>$ among the different data sets. 
The skewness is definitely negative, between -0.052 and -1.75 for Mg$_2$, 
around -0.7 for Mg$b$ and -4. for $<$Mg$_2>$, but shows too much
scatter to be very useful. 

The ADev and ADR are more stable due to their relative insensitivity
to outliers (ADev: 0.0206$\pm$ 0.0014, 0.346 $\pm$ 0.032, 0.02 $\pm$
0.001 and ADR: 1.265 $\pm$ 0.058, 1.105 $\pm$ .12, 1.7 $\pm$ 0.09 for
Mg$_2$, Mg$b$, $<$Mg$_2>$ respectively). So for the rest of our study
we prefer focusing on these two statistics.

\citet{bend93} noted that Mg residuals could be described by a
Gaussian core with extra galaxies at weak Mg strength. The width of
this Gaussian core is also of crucial interest for the ``ancient
formation'' hypothesis because it represents the amount of scatter
imprinted on the Mg-$\sigma$ relation at formation.  In order to find
its width from the various data sets, we compute the number of
galaxies within bins of 0.01 mag in both Mg$_2$ and $<$Mg$_2>$
residual, and within bins of 0.1 \AA\ in Mg$b$.

\begin{figure}
\epsscale{0.8}
\plotone{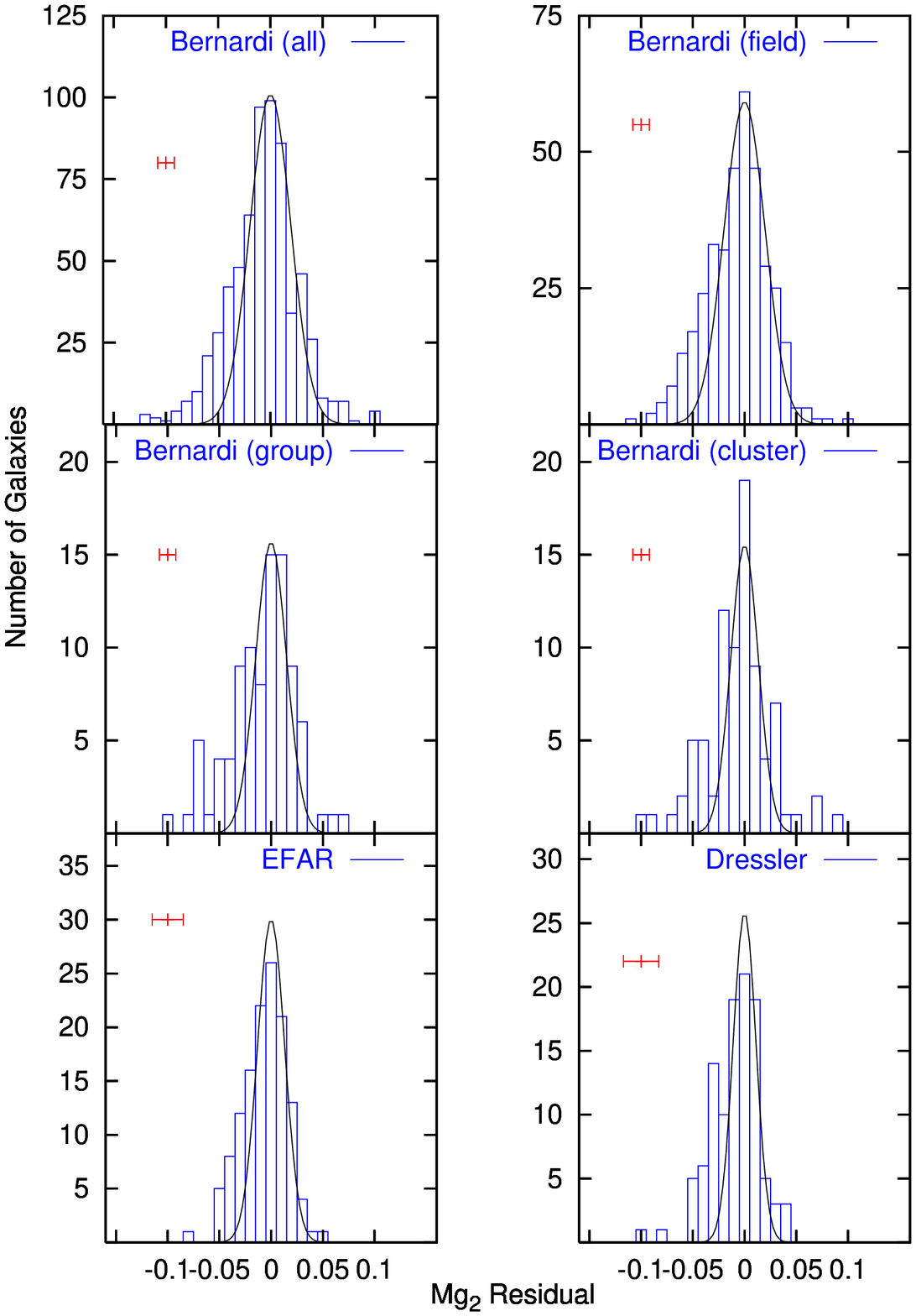}
\figcaption{Number of galaxies versus Mg$_2$ residual. The curve 
is the Gaussian fit for the peak at zero. The bar at top left shows the 
typical uncertainty on the measurements. \label{fig2-a}}
\end{figure}

\begin{figure}
\epsscale{0.8}
\plotone{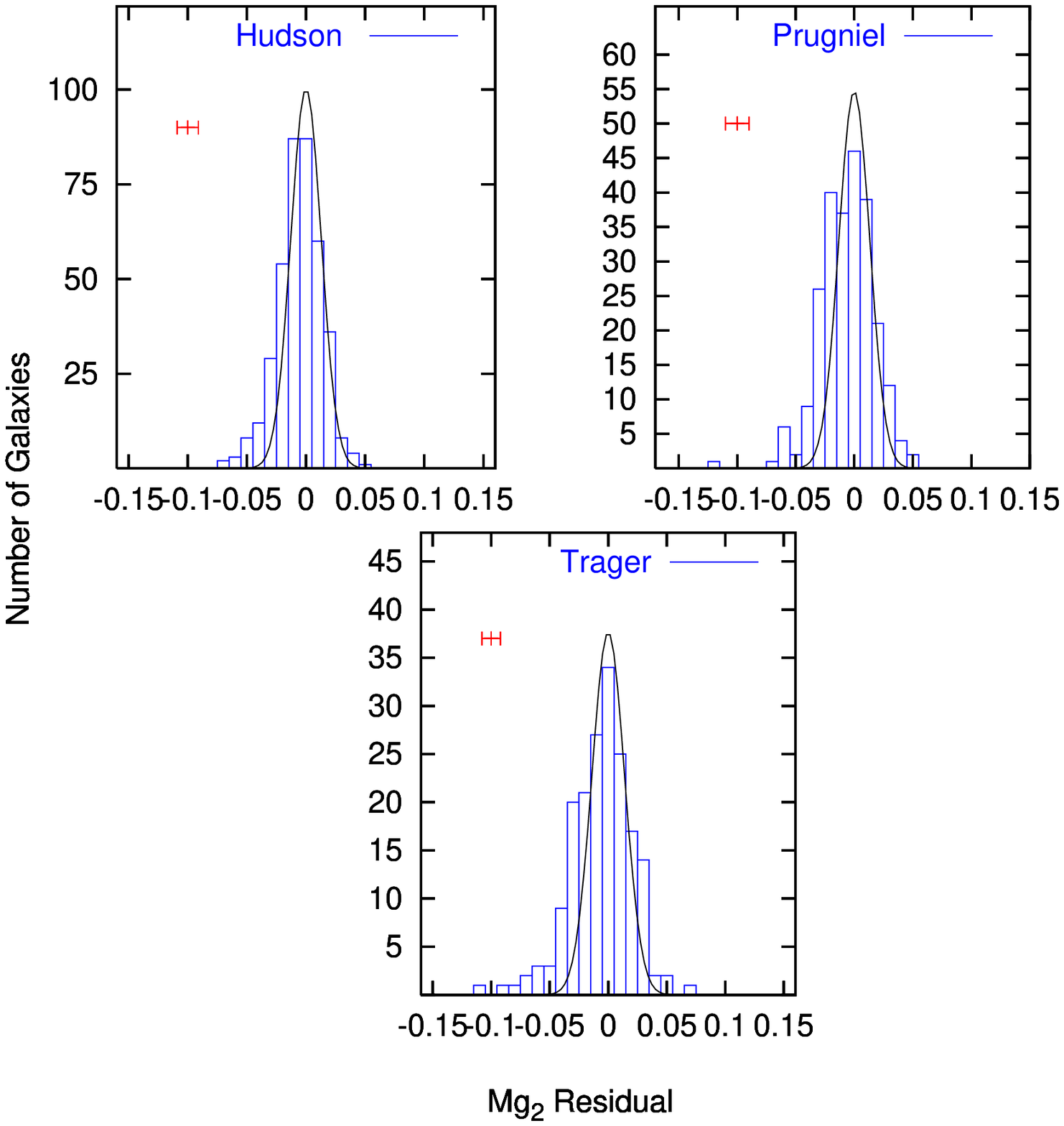}
\figcaption{Number of galaxies versus Mg$_2$ residual. The curve is 
the Gaussian fit for the peak at zero. The bar at top left shows the 
typical uncertainty on the measurements.\label{fig2-b}}
\end{figure}

\begin{figure}
\epsscale{0.8}
\plotone{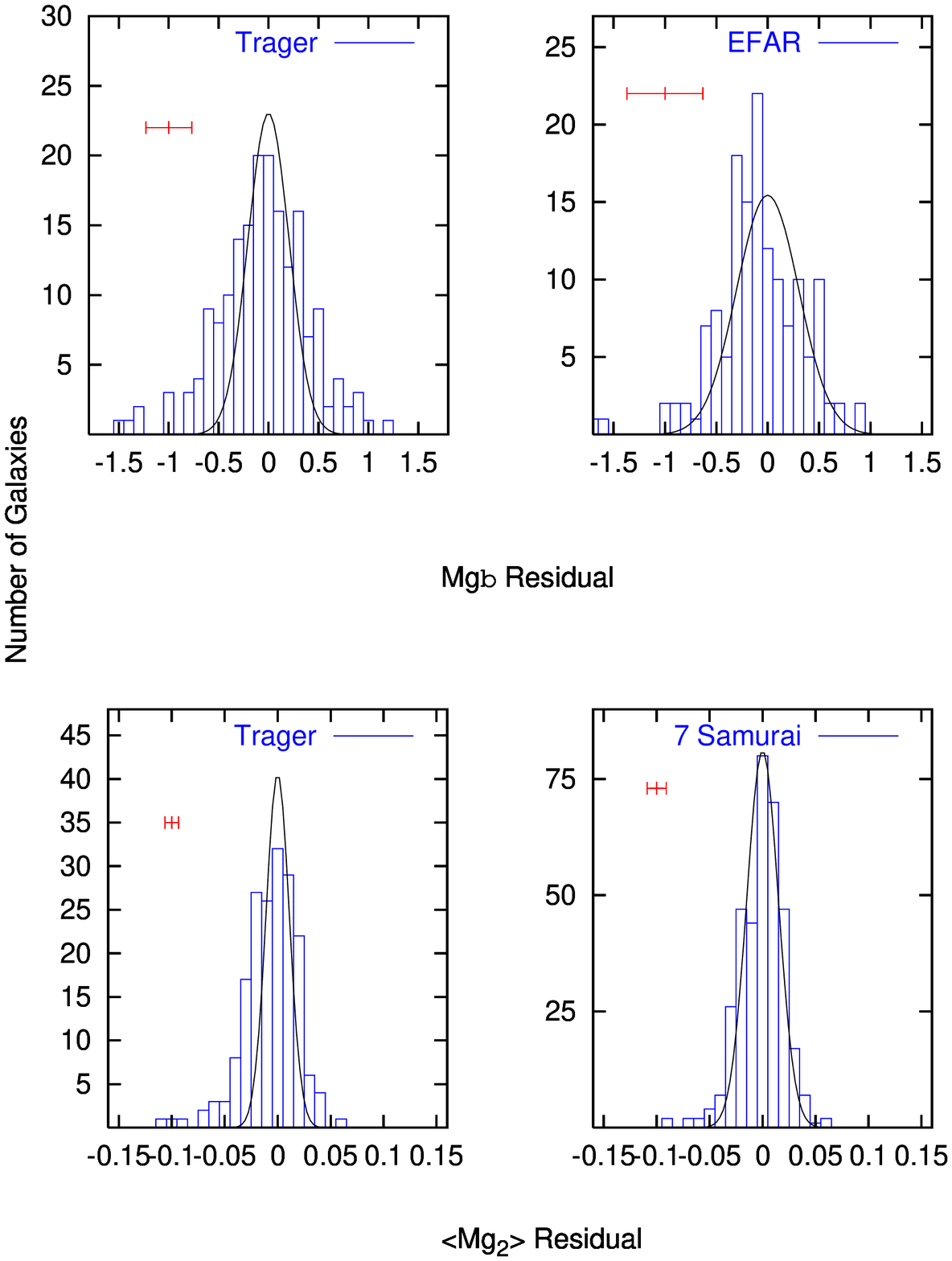}
\figcaption{Number of galaxies versus Mg$b$ (or $<$Mg$_2>$) residual. The 
curve is the Gaussian fit for the peak at zero. The bar at top left shows the 
typical uncertainty on the measurements.\label{fig2-c}}
\end{figure}

We fit with a Gaussian the principal peak from Mg residual distributions, 
weighting by $\sqrt{N}$ in each bin (See fig~\ref{fig2-a},
fig~\ref{fig2-b} and fig~\ref{fig2-c}).  These fits yielded widths
that were very close for all the data sets except for the EFAR and
\citet{dress91} samples. But these are also the samples with the
largest observational errors, and we conclude that this is why the
fitted Gaussians are more broad.  On average, we found $\sigma_g$
equal to 0.0147 $\pm$ 0.002 for Mg$_2$, and 0.013 $\pm$ 0.002 for
$<$Mg$_2>$. Because of the difference of units, $\sigma_g$ for the
Mg$b$ index is broader: $\sigma_g = 0.25 \pm 0.05$.

\placefigure{fig2-a}
\placefigure{fig2-b}
\placefigure{fig2-c}

All data sets give remarkably similar characteristics. We see no
statistically significant difference in width or asymmetry among E and
S0 galaxies in different environments (field, group or cluster). Also
note that our method of line-fitting yields Mg$_{300}$ field-group-cluster
offsets even smaller than those computed by \citet{bernar98} from the
same data.

For the remainder of the paper, we prefer the ADev and ADR statistics
for providing the first-order discriminant for the synthetic
Mg-$\sigma$ diagrams, whose generation discuss in the following section.

\section{Merger Models}

In order to understand the effects of galaxy merging on the scatter in
the Mg-$\sigma$ relation we constructed hybrid models that contain
simple galaxy merging trees and rudimentary chemical evolution but a
relatively rigorous treatment of the effect of stellar ages on Mg
feature strength. The aim of the resultant Monte Carlo simulation is
to make a synthetic Mg-$\sigma$ diagram where the stretch along the
$\sigma$ axis is heuristic and approximate but where the scatter in
the Mg strength direction is modeled well. Also, {\it relative} zeropoint
changes in Mg strength are reliable; for example, the weakening of Mg
feature strength with increasing lookback time is well modeled (if one
can assume that ellipticals today were also ellipticals in the past).

There are several simplifying assumptions in the models. Chemical
enrichment proceeds in a schematic way. A mass-[Mg/H] relation is
assumed, and a pre-merger galaxy fragment is assumed to have the
[Mg/H] appropriate for its mass. Upon merging, newly incorporated
gas is assumed to become stellar with an [Mg/H] derived from the
post-merger mass. Only the gas enriches in this way; the stellar
portion of the merger fragments retain their original abundance. 
 Using this scheme, the Mg abundance is driven toward higher
values in larger galaxies, as is observed. This is one of the more
heuristic elements of the models, the chief intent being to produce a
spread of galaxy Mg strengths. Almost any mechanism that accomplishes
this is sufficient for the purposes of this exercise since it is the
spread of Mg strength from stellar age effects that is under study,
not the origin of the relationship itself.

The choice of [Mg/H] rather than [Fe/H] is mandated by observation. At
high metallicities, elliptical galaxies are seen to have approximately
constant Fe-feature strength as a function of velocity dispersion
\citep{trag98}. Mg-feature strengths increase with velocity dispersion,
and the tight relation between the two variables is the cause of this
investigation. Scaled-solar stellar population models predict that
both species should increase together with increasing metallicity
(e.g. \citet{wor94}), so it is mostly Mg (and presumably other light metals)
that drives the increase of abundance in the largest galaxies. Since
the deep Mg feature at 5100\AA\ is the easily measured, [Mg/H] becomes
our abundance parameter. We choose [Mg/H]$\approx$0.6 for the most
extreme $\sigma$=390 km/s, $M$=10$^{14}$ M$_\odot$ giant
ellipticals, and [Mg/H]$\approx$0 for $\sigma$=150 km/s,
$M$=5.7$\times$10$^{12}$ M$_\odot$ smaller elliptical galaxies.

The relation between velocity dispersion and mass is a projection of
the fundamental plane. Specifically, a relation in Kappa-space from
\citet{bend92} was chosen to pass from
compact ellipticals through giant ellipticals, namely $\kappa_2 = 2.2 +
0.2\kappa_1$. This translates to a rough relation between galaxy mass
and velocity dispersion: $M = 0.0322 \sigma_0^{-3.394}$. This relation
is used in the simulations to translate mass (the fundamental variable
in the simulations) to velocity dispersion. The artificiality of this
relation does not affect the conclusions of this paper, and
its simplicity allows us to explore the stellar population effects
caused by complex merger histories.

The part of the models that is not approximate is the treatment of the
stellar populations. Quick interpolation tables were summarized from
\citet{wor94} models using both the original
Yale/VandenBerg evolution and the \citet{ber94} isochrones
for younger ages. Lookups can be done for $V$ luminosity and the three
Mg indices. The simulation delivers a galaxy formed at several (or
many) epochs. These are collected as a list of ages, masses, and
[Mg/H] abundances. Index strengths are computed as a function of age
and [Mg/H] and weighted by $V$ luminosity computed as a function of
age, [Mg/H], and mass for each tagged star parcel within the final
galaxy assemblage.

The main parameters in the models are four. The number of equal-mass
merger fragments that will eventually combine to form the final
galaxy, the fraction of the final galaxy that formed at the redshift
of formation $z=z_f$, the gas fraction of the mergers $F_{\rm gas} = m_{\rm
gas}/m_{\rm total}$, and a probablility envelope that is either flat
with time between $z_f$ and $z=0$ or is proportional to $(1+z)^n$,
where $n$ is a free parameter.

Other parameters control cosmology, set $z_f$, set the number of
galaxies to be generated and their mass range, introduce artificial
scatter into the computed Mg-$\sigma$ relation, and provide for extra
(or not as much) Mg enrichment at each star formation episode.

Throughout this paper we adopt $H_0 = 60\, km/s/Mpc$, $\Omega = .2$,
$\Omega_R = .8$, $\Omega_{\Lambda} = 0$. With these parameters, the
universe is 13.8 Gyr old, and $z_f = 5$ corresponds to a lookback time
of 12.3 Gyr.
We flagged galaxies that experienced star formation
within the last 50 Myr as ``star forming'' and they are omitted from
all statistical calculations.
As the majority of data sets feature the Mg$_2$ index, we use
this index for all of the following discussion even though the models
predict all index varieties. 

\section{Merger Model Discussion}
\subsection{Basic Model Behavior}

\begin{figure}
\epsscale{0.32}
\plotone{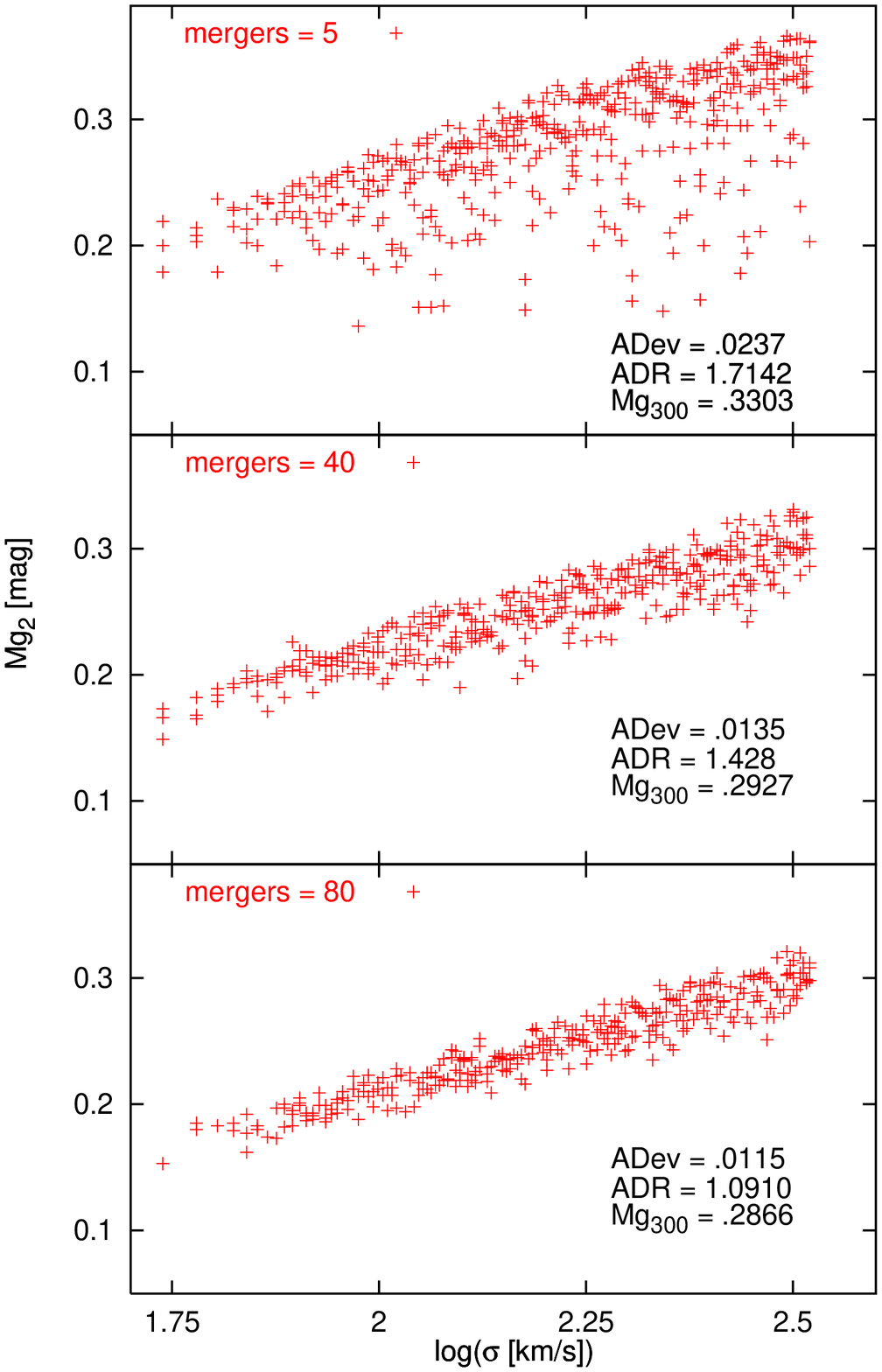}
\figcaption{Influence of the number of equal-mass merger for 5, 40 
 and 80 mergers, 
with a uniform probability envelope,  F$_{gas} = 0.2$, and a fraction of 
galaxy formed at z=z$_f$ of F$_{ini} = 0$. \label{fig3-a}}
\epsscale{0.32}
\plotone{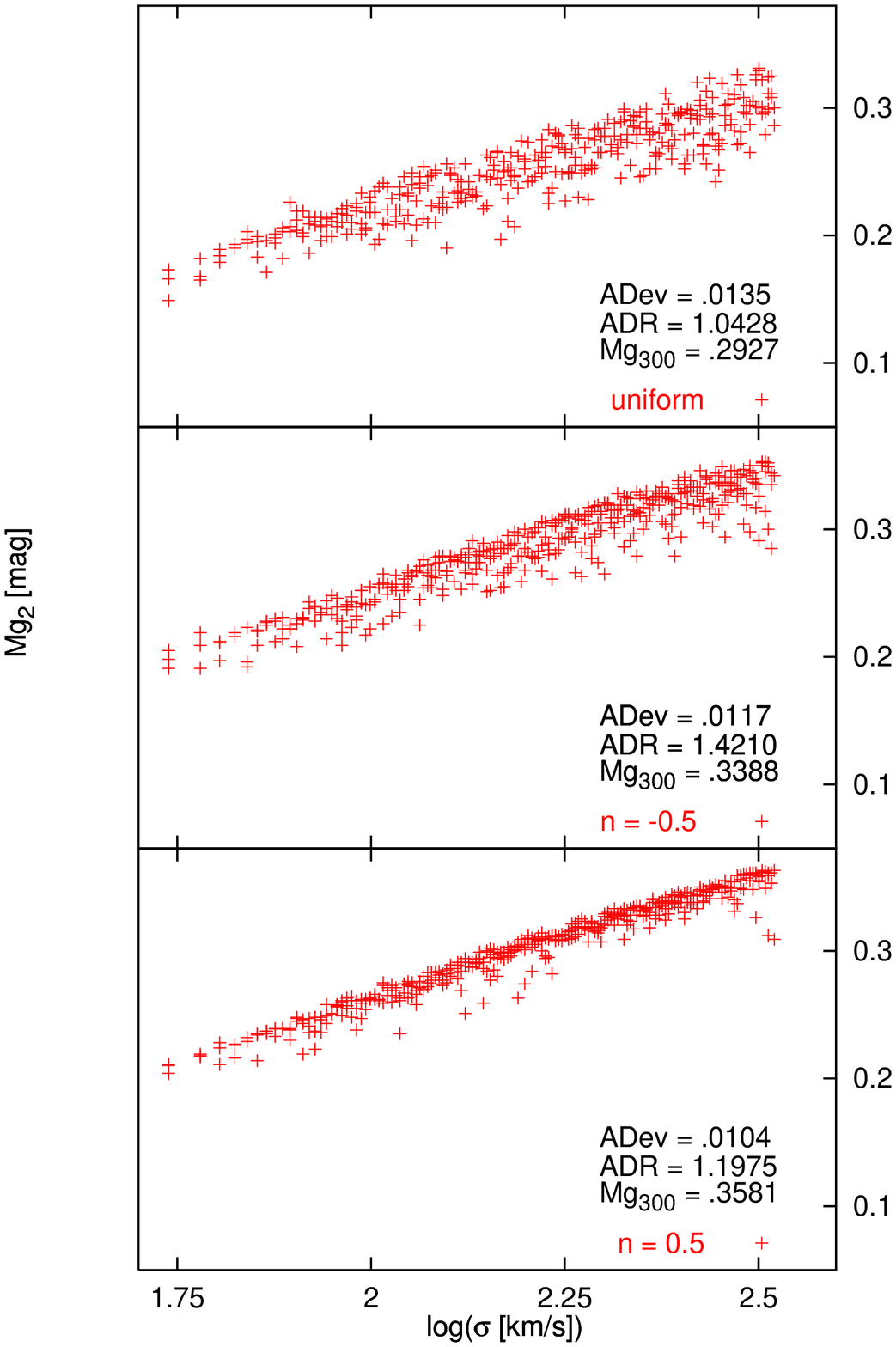}
\figcaption{Influence of the probability for uniform probability,
 $n=-0.5$, and $n=0.5$, with 40 mergers,
 F$_{gas} = 0.2$, and F$_{ini} = 0$ .\label{fig3-b}}
\end{figure}

As an introduction to the model output, we focus on examples of the
effect of our four main parameters.  Figure~\ref{fig3-a} shows the
effect of varying the number of equal-mass mergers that eventually
form the galaxy. With a lot of mergers (80 mergers, for example), the
asymmetry becomes low since all the galaxies have almost the same
history. That is, they all suffer a peppering of small amounts of star
formation over the age of the universe.  But with very few mergers (5
mergers, for example) the dispersion is larger and the asymmetry is
much more pronounced. Each galaxy experiences a distinct timing
fingerprint for mergers and thus are more individualistic, some having
had recent star formation, others having had none.

\begin{figure}
\epsscale{0.32}
\plotone{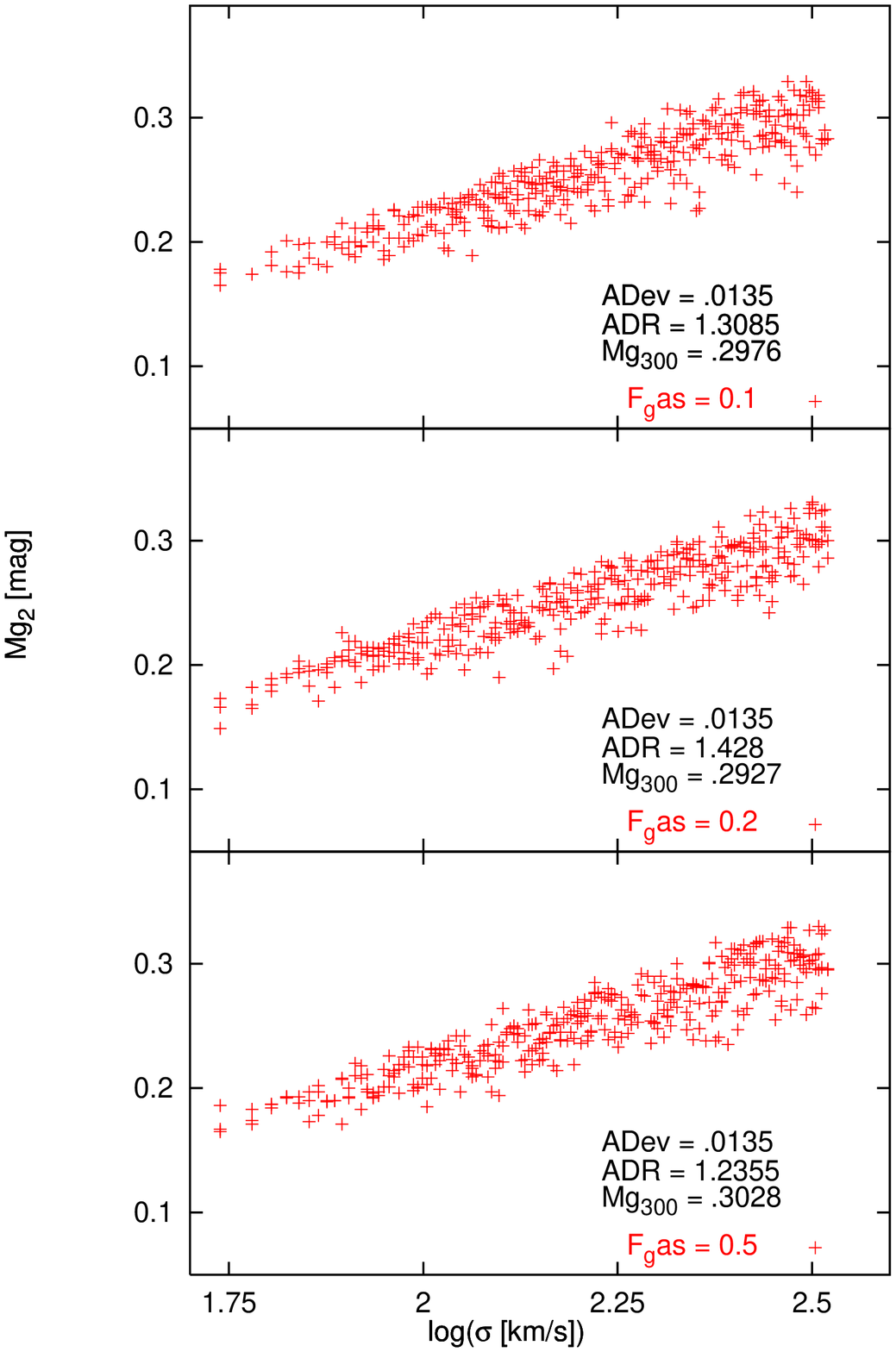}
\figcaption{Influence of the fraction of gas in each 
merging event, for 3 different gas fractions, 
F$_{gas} =$ : .1, .2 and .5, for 40 mergers, a uniform probability 
envelope, and a fraction of galaxy formed
at z=z$_f$, F$_{ini} = 0$.\label{fig3-c}}
\epsscale{0.32}
\plotone{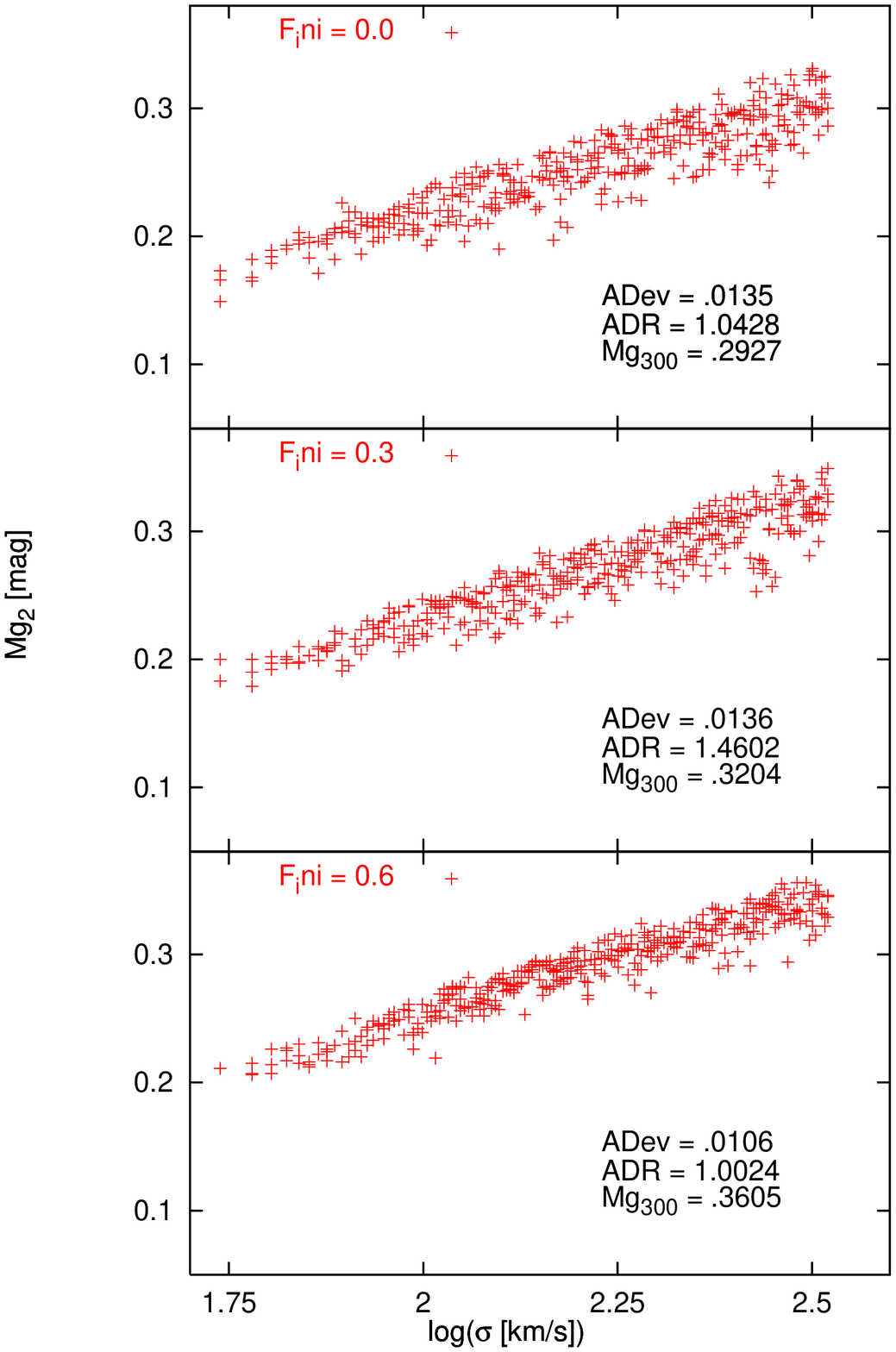}
\figcaption{Influence of the fraction of the galaxy formed at z=z$_f$, for 3
different initial fractions, F$_{ini} =$ 0, 0.3 and 0.6, for 40 
mergers, a uniform probability envelope and F$_{gas} = 0.2$.  
\label{fig3-d}}
\end{figure}

\begin{figure}
\epsscale{0.8} 
\plotone{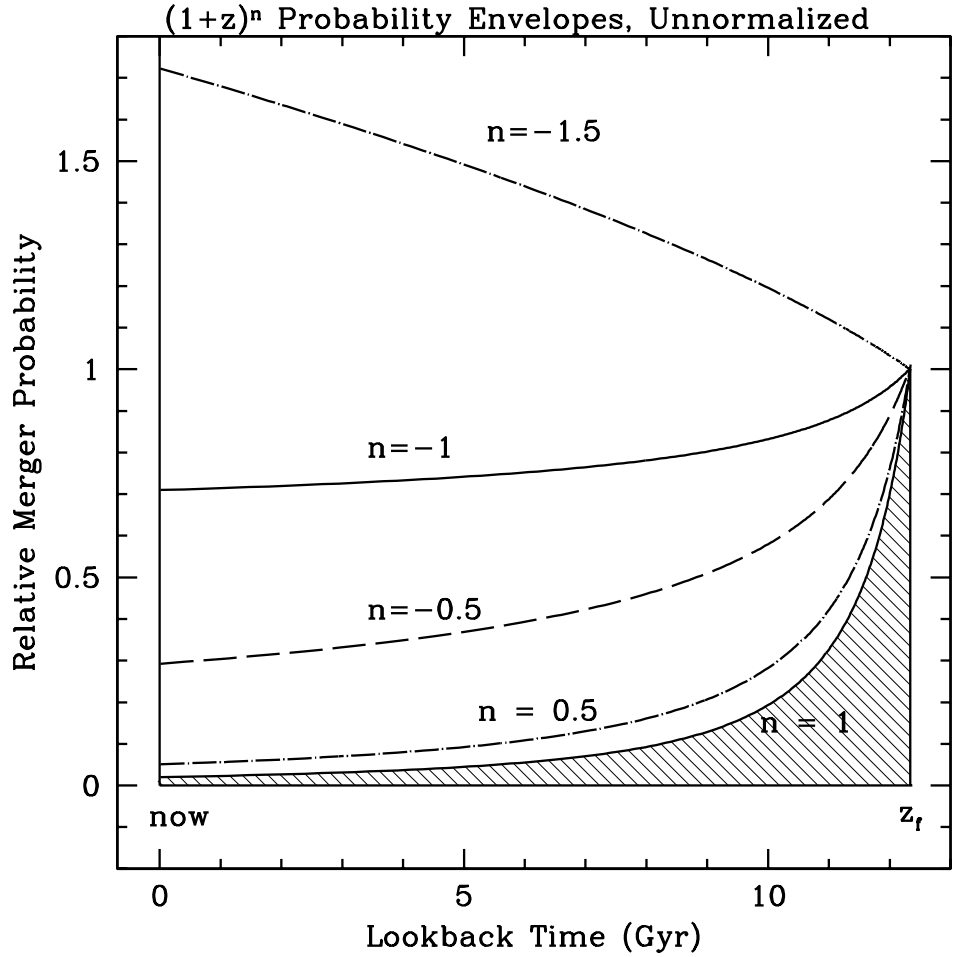} 
\figcaption{Merger probability as a
function of time. For higher $(1+z)^n$ exponents, merging events
happen preferentially early. The ``uniform'' probability often used in
this article would be represented by a horizontal line in this
diagram.\label{fig4}}
\end{figure}

\placefigure{fig3-a}
\placefigure{fig3-b}
\placefigure{fig3-c}
\placefigure{fig3-d}
\placefigure{fig4}

The shape of the probability envelope has a strong impact as well. The
frequency of mergers is often parameterized in the literature as being
proportional to $(1+z)^n$, where $n$ is a free parameter.
Figure~\ref{fig4} translates some of these choices from redshift to
time.  One can see that $n = -1.2$ roughly corresponds to a uniform
probability with time (note that when we say ``uniform probability''
throughout this paper, we mean a perfectly flat [with time] curve,
avoiding altogether the $(1+z)^n$ formalism.); $n = -0.7$ corresponds
to a present-day merger frequency about half of that at $z=z_f$, and
$n > 0$ corresponds to merger rates strongly skewed to high
redshift. These numbers change somewhat with cosmology. 

As fig~\ref{fig3-b} shows, it is plain that
with $n = 0.5$ the obtained distribution is very thin.
Since mergers are concentrated in the past, few are lucky enough to
have had recent star formation. The relation thickens with decreasing
$n$. 

The parameter $F_{gas}$ is the fraction of gas used in each merger event.
In Figure~\ref{fig3-c} we see that more gas leads to more star
formation and hence more young galaxies. Less gas leads to weak
starbursts and a narrow Mg-$\sigma$ relation.

The fraction of galaxy formed at $z=z_f$, $F_{ini}$, has less of a
dramatic effect (see fig~\ref{fig3-d}) except for the trivial case of
$F_{ini}\approx 1$. There is little difference between forming half of
the galaxy at $z_f$ and forming the galaxy entirely by mergers, rather
to the initial surprise of the authors. Apparently, it is the last few
Gyr of history that are the most important.

\subsection{Maximum Merger Models}

We now search quantitatively for solutions under the maximum merger
hypothesis.  For fair statistical comparison, the artificial data were
further randomized by an artificial Gaussian scatter $\sigma_a = 0.01$
mag, representing the observational error only. That is, $\sigma_a <
\sigma_g$, the width of the Gaussian core of the residual
distribution. This assumption is modified for the ``ancient
formation'' hypothesis described below.

Because we compare to the mean of all of the observational data, we
analysed data from a similar number of artificial galaxies: roughly
2000 Monte Carlo galaxies. The scatter from limited sample size of
course decreases with increased number of samples: for 250 galaxies
the errors were .0009 and .2 for respectively the ADev and the ADR
(computed from multiple realizations of the same simulation). With
2000 galaxies we obtained .0002 for the ADev and .04 for the
ADR. These errors are not too different from those seen in the
observational data sets, so it is all but certain that sampling error
causes most of the statistical deviations between data sets. This
leads to important guidelines for future work at higher redshift:
large sample size is crucial if the scatter is to be adequately
characterized, and errors may not creep too much higher than their
nearby brethren.

\begin{figure}
\epsscale{0.7}
\plotone{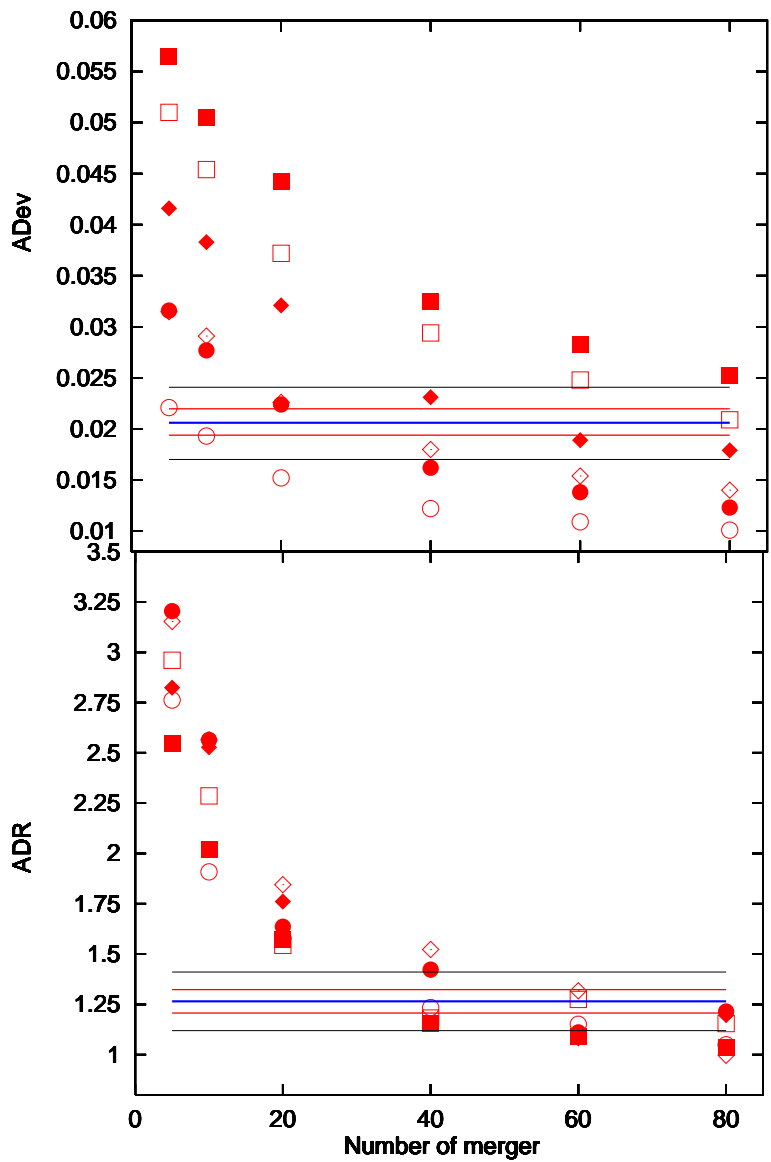}
\figcaption{
ADev (top) and ADR (bottom) for a uniform probability versus
the number of mergers. For F$_{ini} = 0.0$, filled bullets, diamonds, and 
squares represent F$_{gas} = .1, .2, .5$, respectively. For F$_{ini} = 
0.5$, open bullets, diamonds, and squares correspond to F$_{gas} = .1, 
.2, .5$, respectively. The central lines are the observational means 
with other lines situated at $\pm \sigma$ and 
$\pm 2.5\, \sigma$. An artificial observed scatter of
$\sigma_a = 0.01\,mag$ was included.
\label{fig5-a}}
\end{figure}

\begin{figure}
\epsscale{0.7}
\plotone{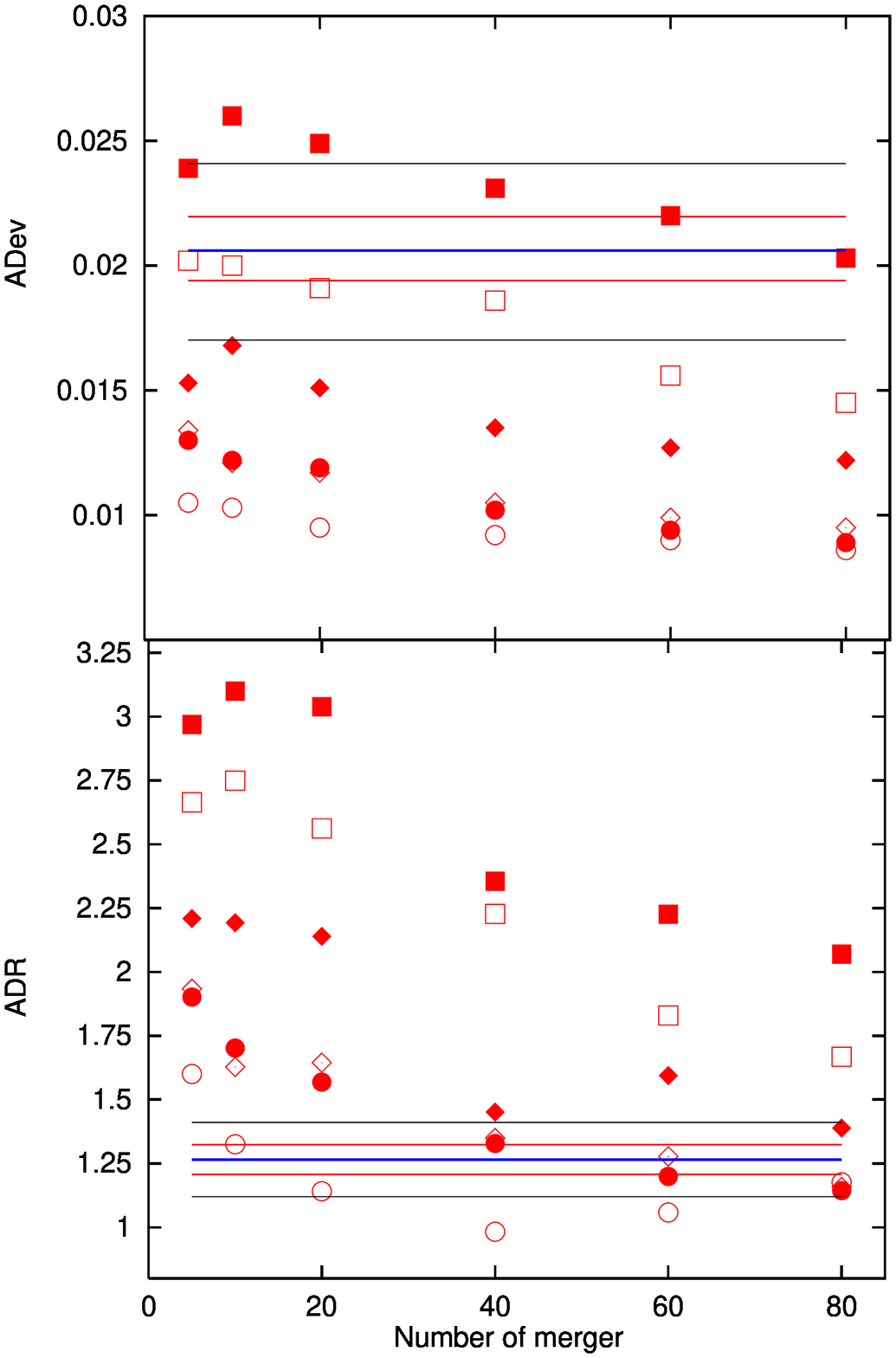}
\figcaption{
ADev (top) and ADR (bottom) for a (1+z)$^0$ probability versus
the number of mergers.  An artificial observed scatter of
$\sigma_a = 0.01\,mag$ was included. Symbols and lines are the same as 
fig.~\ref{fig5-a}.
\label{fig5-b}}
\end{figure}

\begin{figure}
\epsscale{0.7}
\plotone{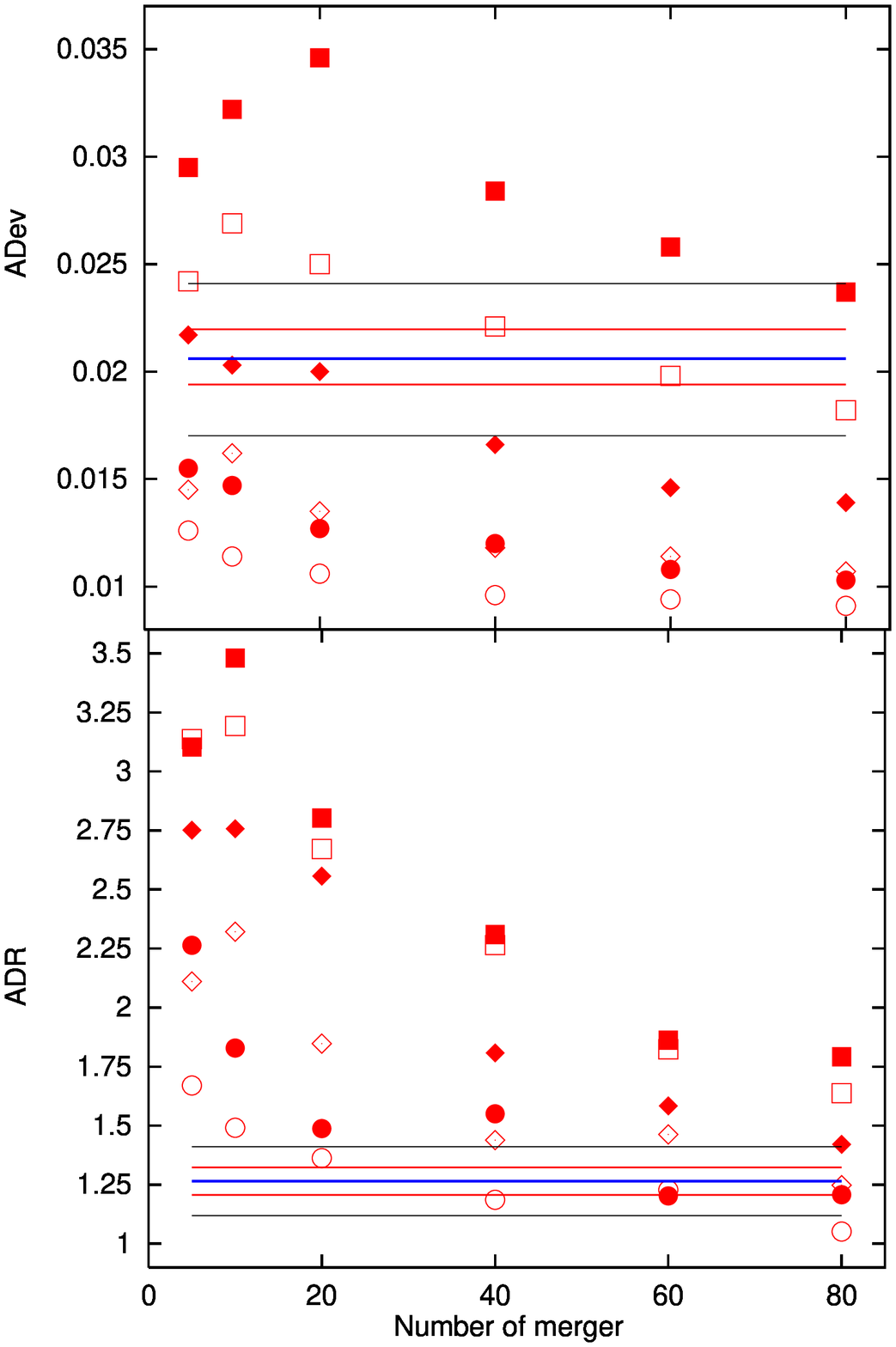}
\figcaption{
ADev (top) and ADR (bottom) for a (1+z)$^{-0.5}$ probability versus
the number of mergers. An artificial observed scatter of
$\sigma_a = 0.01\,mag$ was included. Symbols and lines are the same as 
fig.~\ref{fig5-a}.
\label{fig5-c}}
\end{figure}

\placefigure{fig5-a}
\placefigure{fig5-b}
\placefigure{fig5-c}

Figures \ref{fig5-a} through \ref{fig5-c} show our main diagnostic
statistics, ADev and ADR, for models that cover the grid of input
parameters. Data for a uniform probability are shown in
Figure~\ref{fig5-a}, for a $(1+z)^0$ envelope in Figure~\ref{fig5-b},
and for a $(1+z)^{-0.5}$ envelope in Figure~\ref{fig5-c}.  Different
$F_{ini}$ and $F_{gas}$ possibilities are plotted as various symbol
types as a function of the number of mergers. The trend with
increasing number of mergers is clear: a tighter, more symmetrical
relation is found. If this seems counterintuitive because with many
mergers one should expect a more volatile and bursty collection of
galaxies, consider instead that each galaxy has a more similar history
if the number of mergers grows large. See also Figure \ref{fig3-a}.
We plot a good sampling of the possibilities, representing F$_{ini} =
0.0$ or $0.5$ and F$_{gas} = .1, .2, .5$ with different symbol
types. To compare directly with the observational data, we plot the
mean observational values found in $\S2$ with $\pm 1.0\,\sigma$ and
$\pm 2.5\,\sigma$ errors.

Other trends are also apparent. At a
given F$_{ini}$, the ADev increases with the gas fraction since with
more gas available more stars will form, making the mean stellar
population more volatile. Similarly, at fixed gas fraction, the ADev
decreases with higher $F_{ini}$. Indeed, if $F_{ini}$ is set to one,
the relation becomes razor thin since the whole galaxy formed at $z=z_f$.
Except for values approaching one, the authors were surprised at the
ineffectiveness of $F_{ini}$ to modify the Mg-$\sigma$ relation by
much; it is the weakest of the four parameters.

As for the influence of when in time the mergers are likely to happen,
we compare the three figures. It is rather difficult to find many
points that land simultaneously in the 2.5-$\sigma$ regions of both
ADev and ADR statistics, but to the extent that some do, most of them
are in the ``uniform probability'' figure, around 40 mergers. Any
probability exponent $n$ greater than zero yields Mg-$\sigma$
relations that are even thinner than the $n=0$ case, and we are forced
to conclude that elliptical galaxy formation that was concentrated to the first
half of a Hubble time is ruled out under the maximum merger hypothesis.

\subsection{Ancient Formation Models}

We now relax our assumption that the Mg-$\sigma$ relation is
inherently thin. We suppose instead that there is a intrinsic,
symmetric scatter $\sigma_g$ equal to that which is observed today,
but imprinted at formation. The mean width derived in $\S$2 is
$\sigma_g = 0.015$ mag. This is rather narrow and not too different
from the observational scatter, 0.010 mag. In fact, consideration of
these numbers implies that the true galaxy-intrinsic Gaussian-random
scatter is around 0.01 mag. It will be interesting to see if future
high-precision data upholds this conclusion.

\begin{figure}
\epsscale{0.7}
\plotone{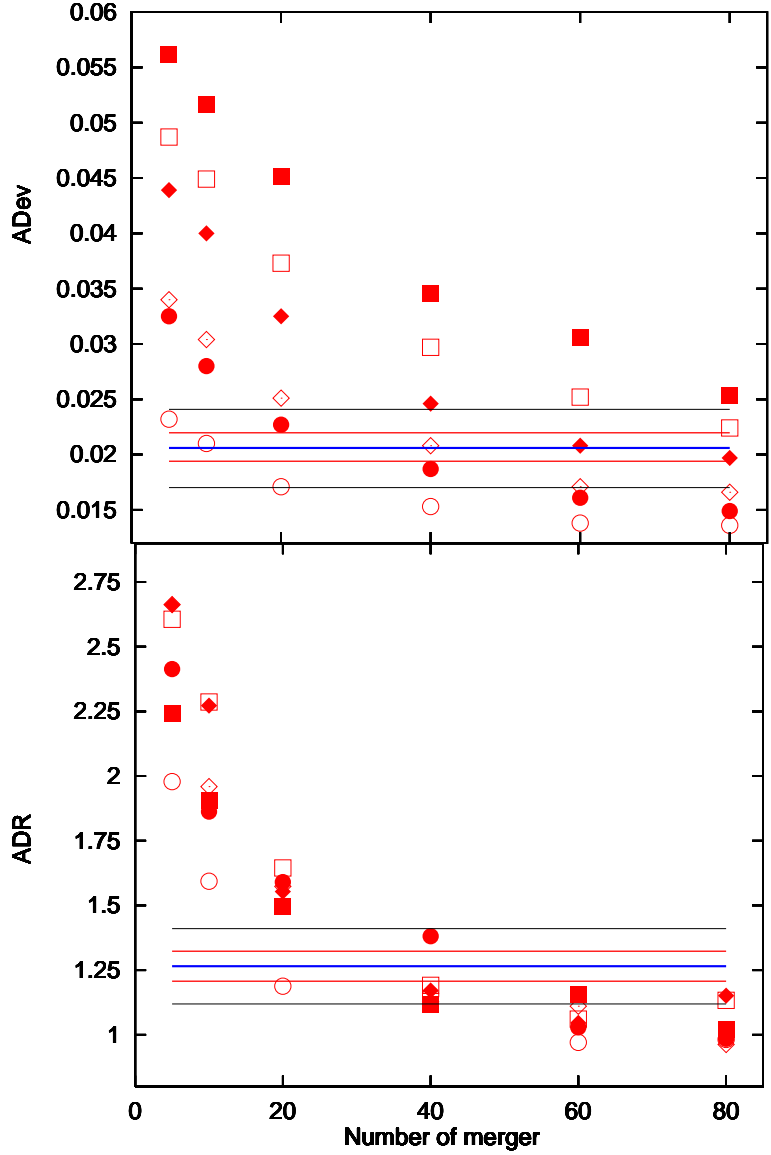}
\figcaption{ADev (top) and ADR (bottom) for a uniform probability  
versus the number of mergers, with an artificial scatter 
$\sigma_g = 0.015\, mag$. For F$_{ini} = 0.0$, filled bullets, diamonds, and 
squares represent F$_{gas} =$ .1, .2, .5, respectively. For F$_{ini} = 
0.5$, the open bullets, diamonds, squares correspond to F$_{gas} =$ .1,  
.2, .5, respectively. 
The central lines are the observational means with other lines situated 
at $\pm \sigma$ and $\pm2.5\, \sigma$ . \label{fig6-a}}
\end{figure}

\begin{figure}
\epsscale{0.7}
\plotone{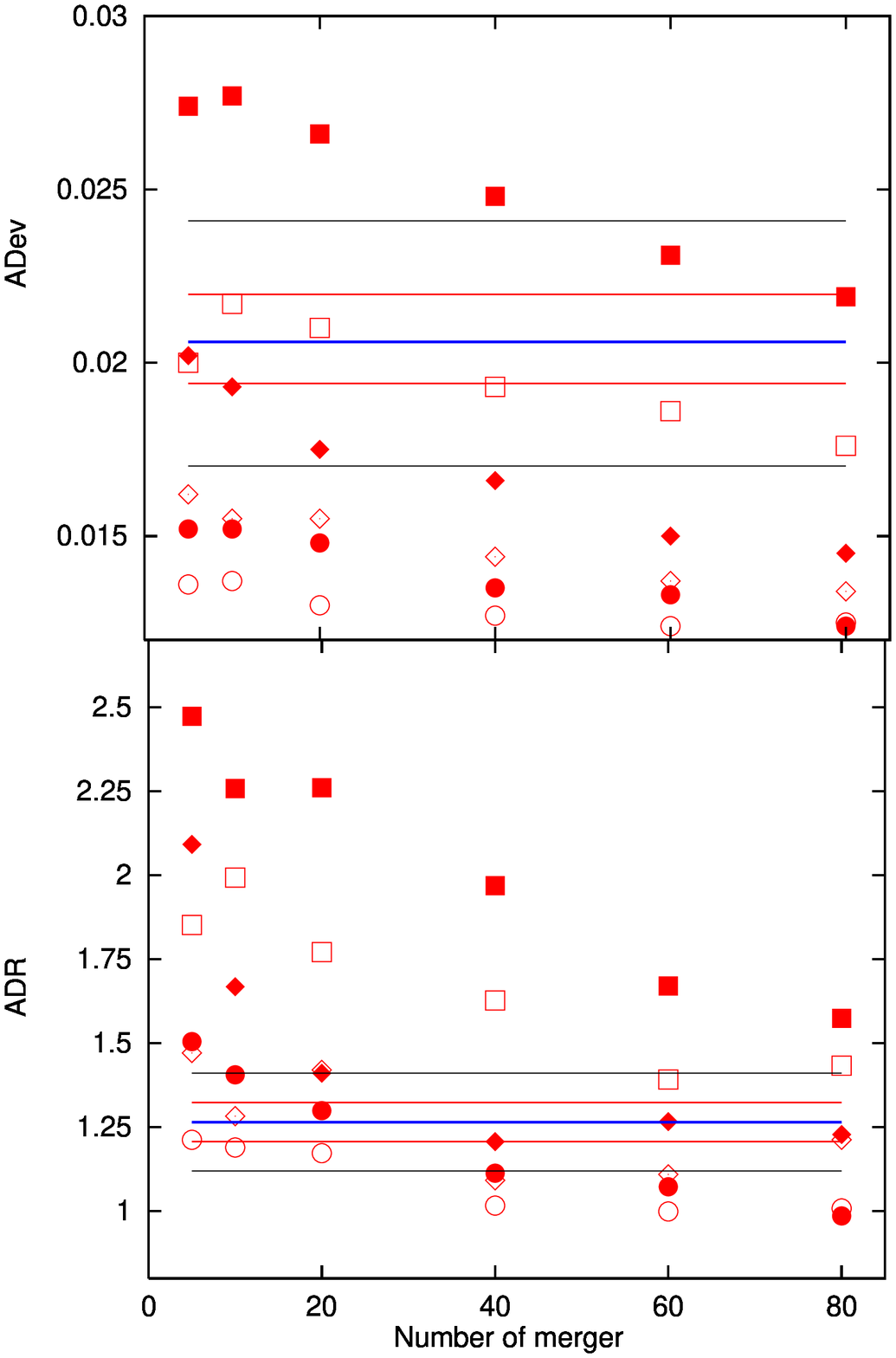}
\figcaption{ADev (top) and ADR (bottom) for a (1+z)$^0$ probability  
versus the number of mergers, with an artificial scatter 
$\sigma_g = 0.015\,mag$. Symbols and lines are the same as 
fig.~\ref{fig6-a}. \label{fig6-b}}
\end{figure}

\begin{figure}
\epsscale{0.7}
\plotone{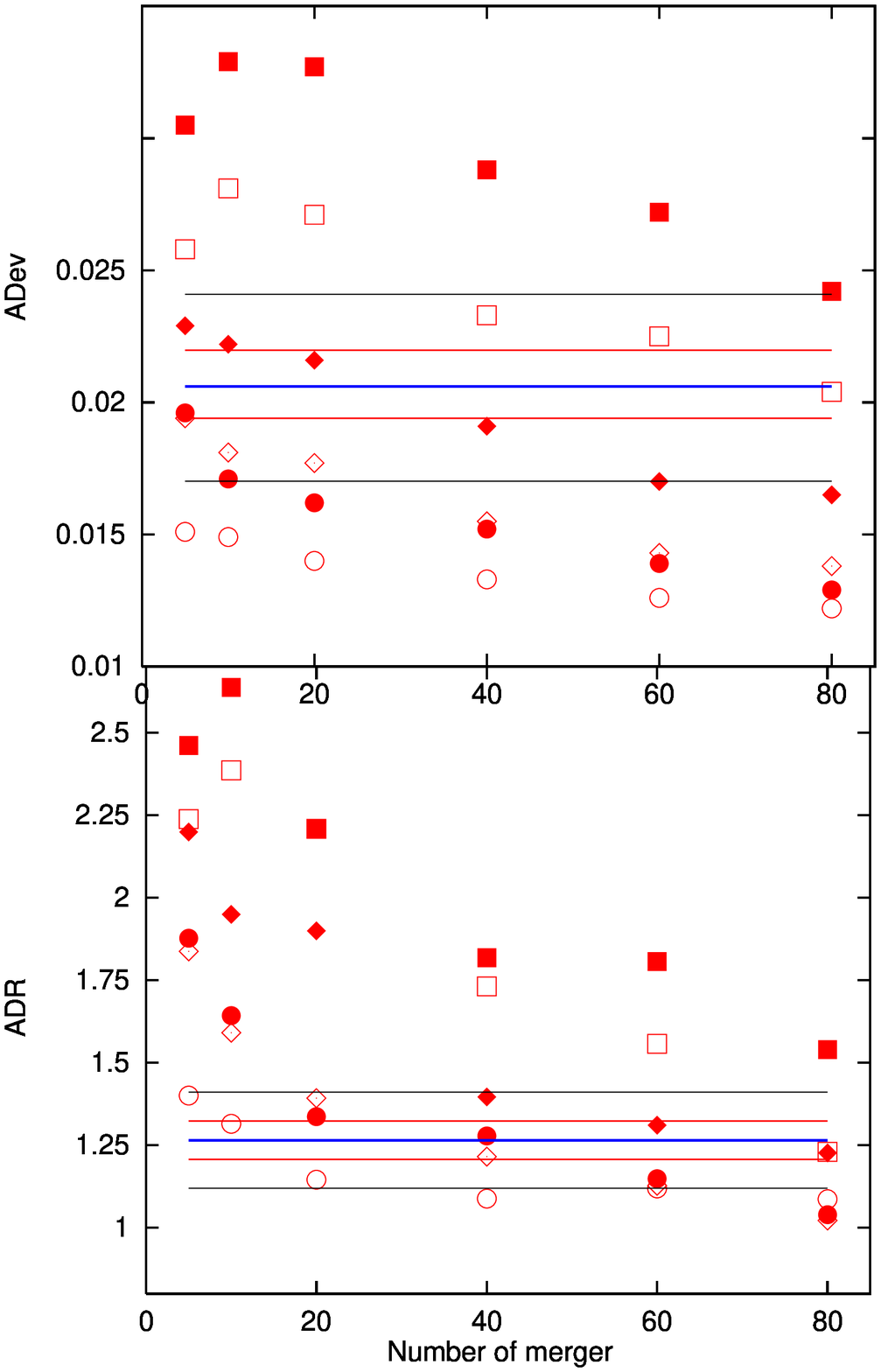}
\figcaption{ADev (top) and ADR (bottom) for a (1+z)$^{-0.5}$ probability  
versus the number of mergers, with an artificial scatter 
$\sigma_g = 0.015\,mag$. Symbols and lines are the same as 
fig.~\ref{fig6-a}.  \label{fig6-c}}
\end{figure}

Under the ancient formation hypothesis, only the asymmetric tail of
the distribution is produced by late star formation. Do these models
fit better? We plot the three different probability cases exactly as
in the previous section: uniform probability (fig~\ref{fig6-a}),
$(1+z)^0$ envelope (fig~\ref{fig6-b}) and $(1+z)^{-0.5}$ envelope
(fig~\ref{fig6-c}). We derived the statistics using 2000 galaxies, and
show the possibilities allowed by $F_{gas} = 0.1,0.2,0.5$ and $F_{ini}
= 0.0$ or $0.5$.  

\placefigure{fig6-a}
\placefigure{fig6-b}
\placefigure{fig6-c}

We find that most of the remarks from last section are still
valid. The cases with 5 or 10 mergers do not fit, and the most
numerous solutions still occur for uniform probability and $N_{\rm
merger}=40$.  The main difference is that the additional artificial
scatter calms the asymmetry somewhat, and this allows a few more
successful models. We even find a couple of models in the $n=0$ panel
that land within the error bars for both statistics, although mostly we find
the width of the Mg-$\sigma$ relation too narrow.

The astounding similarity between maximum-merger and minimum-merger
cases is driven by the high quality and consistency of the
observational data. With the low and high bounds of allowed artificial
scatter set at 0.010 and 0.015 mag, tight reigns are held on our
freedom to interpret the results.

\subsection{Constraints from Mg$_{300}$}

To this point in the paper we have examined only 'primary' statistics
ADev and ADR for the width and asymmetry of the Mg-$\sigma$
residuals. We can explore the situation a little more by also
considering the absolute Mg strength of the Mg-$\sigma$ relation,
parameterized by Mg$_{300}$, the median Mg strength at $\sigma =$ 300
km/s. Including Mg$_{300}$ along with ADR and ADev, we define a
goodness parameter to evaluate the relative merit of the various
models:
\begin{displaymath} 
\sigma^2_G = \frac{(ADev_i-\overline{ADev_{obs}})^2}{\sigma^2_{ADev}} +
 \frac{(ADR_i-\overline{ADR_{obs}})^2}{\sigma^2_{ADR}} +
\frac{(Mg_{300_i}-\overline{Mg_{300_{obs}}})^2}{\sigma^2_{Mg_{300}}}
\end{displaymath}

We took $\overline{ADev_{obs}} = .0206$ , $\overline{ADR_{obs}} =
1.265 $ and $\overline{Mg_{300_{obs}}} = 0.329$, for the different
scatters we had: $\sigma_{ADev} = 1.4\,10^{-3} $, $\sigma_{ADR}=
5.8\,10^{-2} $ and $\sigma_{Mg_{300}} = 0.04$.  This last value is
less well-motivated than the others since we expect considerable
uncertainty in model Mg zeropoint so that cannot adopt the observed
uncertainty for $\sigma_{Mg_{300}}$. On the other hand, 
the overall Mg level is an important diagnostic and too-weak overall
Mg is clearly to be discouraged, so our value of 0.04 is an
approximate compromise.

\begin{figure}
\epsscale{0.7} 
\plotone{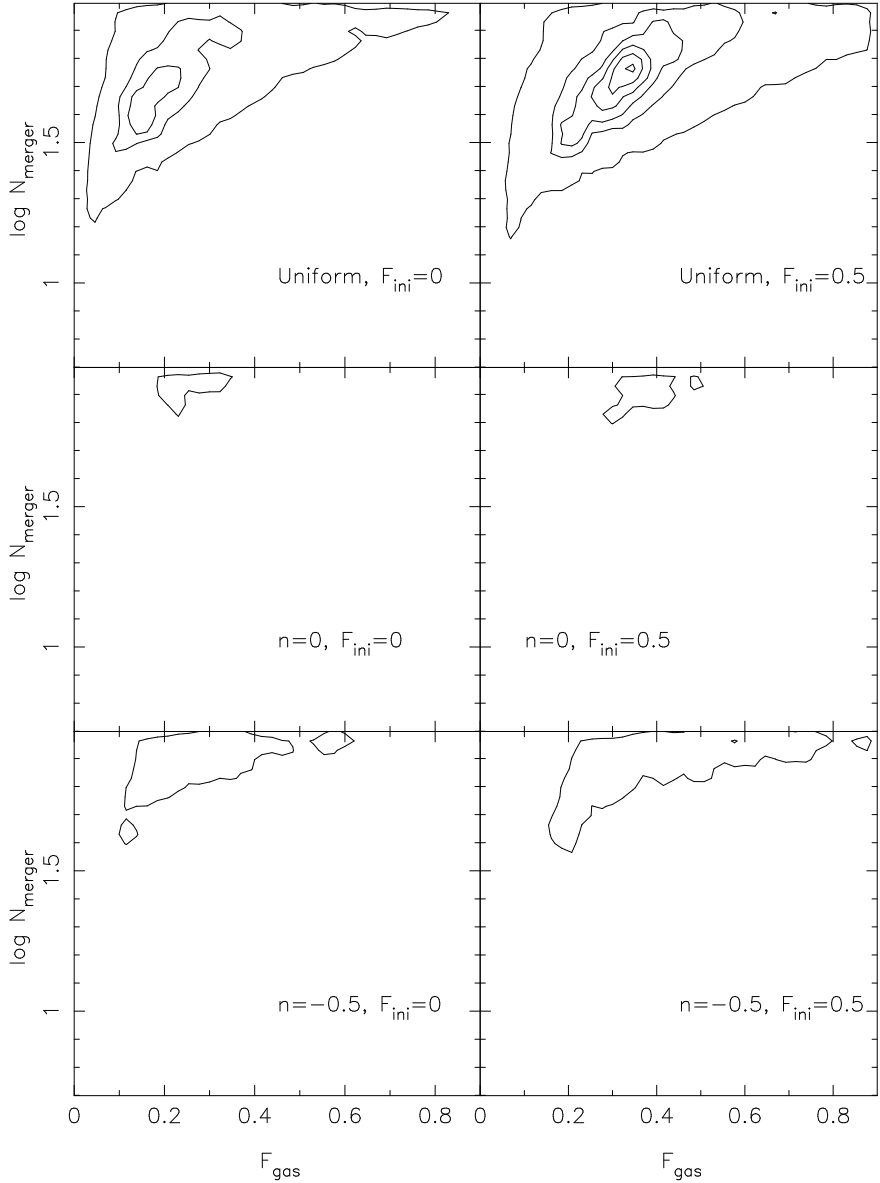} \figcaption{Isovalue contours of
$\frac{1}{\sigma_G}$ for a uniform probability (top), $n=0$
probability (middle), and $n=-0.5$ probability (bottom) for F$_{ini}=
0$ (left) and $0.5$ (right). An artificial observed scatter of
$\sigma_a = 0.01\,mag$ was included. The contours are set at multiples
of $\sqrt{3}/4$ so that borderline acceptability corresponds
approximately to the second contour level, ``good'' fits to the
fourth. \label{fig7}}
\end{figure}

We compute this $\sigma_G$ for the usual 3 cases : uniform merger
probability envelope, (1+z)$^0$ and (1+z)$^{-0.5}$ probability, and
for both F$_{ini} = 0$ or $0.5$. For each case, we make plots from 5
to 100 mergers, and from F$_{gas} = 0$ to F$_{gas} = 0.9$, for 40
values using a linear distribution for the fraction of gas and a
logarithmic distribution for the number of mergers. We
plot isovalues in 1/$\sigma_G$ so that the best models show as maxima
rather than minima. Fig~\ref{fig7}
shows the results for uniform, (1+z)$^0$
and (1+z)$^{-0.5}$ probability envelopes.

\placefigure{fig7}

In this plot there are elongated regions of high $\sigma_G$ values. They
tend to slant, indicating significant degeneracy between gas fraction and
number of mergers in determining a successful model. The amplitude of the
band decreases with the $n$ of the envelope probability.
We see that the valid possibilities shift to slightly higher
values of gas fraction when F$_{ini} = 0.5$.
The uniform envelope probability appears to be the best fit.

\subsection{Constraints from Mean Age and Abundance Spread}

We now check to see if the possibly successful models indicated by
ADev, ADR, and Mg$_{300}$ in Figure \ref{fig7} are compatible with the
abundance distribution and the V-band light-weighted age observed in
nearby galaxies. The abundance distribution is observed to be narrow
in all galaxies the size of M32 and larger \citep{wor96}, with a FWHM
of less than 0.4 dex \citep{grill96} in M32 itself, and about the same
in the Milky Way (e.g. \citet{rana91}). We can check the abundance
distributions of the simulations as well. The chemical evolution of
these models is primitive, but if the abundance distributions are
wider than about 0.5 dex FWHM then we must regard them with
suspicion. Also, if the light-weighted mean age (the closest our
present models can come to approximating a spectroscopic Balmer-metal
feature mean age) falls below about 7 Gyr (the approximate median of
such studies, e.g. \citet{terle02, gonza93}) we likewise regard this
model with much suspicion. Older ages are allowed because the age
zeropoint of the Balmer-metal feature technique is still quite
uncertain.

What we find is emphatic. For $n=0$ or $n=-0.5$ models, no model
passes the ADR/ADev test very well, but they also tend to have an
abundance scatter greater than 0.5 dex. For uniform probability, the
mean galaxy age begins to becomes too young for a combination of high
gas fraction and large numbers of mergers, but this never became a
critical problem because the ADR/ADev test excludes these models as
well. The best surviving model was with uniform probability,
$F_{ini}=0.5$, $F_{gas}=0.35$, and 50 mergers, with an abundance
histogram at the 0.5 dex width limit, mean age of 8.3 Gyr, and ADR and
ADev within the error bars. At large number of mergers, the
$F_{ini}=0.5$ models tend to have a bimodal abundance distribution
caused by the initial metallicity of the galaxy, to which is added
additional material from the gaseous portion of the merger, plus a
low-metallicity peak from the stellar component of the relatively
low-mass merger fragments. In fact, all of the $F_{ini}=0.5$ high
points in Fig. \ref{fig7} are at the borderline of having a too-wide
abundance distribution, and we should penalize these models a couple
of contour levels for this near-transgression. The $F_{ini}=0$ models
do not suffer from this: they all have suitably narrow abundance
distributions. We claim roughly equal success for both
versions of the uniform-probability models, but resounding failure for
models in which galaxy mergers happen preferentially in the past.

\subsection{Model Variants}

Being left with only one fully successful model after trying so many
hundreds left the authors scratching their heads. We decided to
explore other options. Our first option was to abandon the constant
merger mass assumption by adopting a cold dark matter (CDM) power
spectrum for galaxy merger fragments.  We use the formulae from
\citet{white91} and \citet{kauf92} which are based on \citet{press74}
formalism. CDM clusters hierarchically with preferentially small
masses in the early universe, building toward larger structures later
on. We built a lookup table with 40 steps in redshift and 200 steps in
mass in order to quickly invert the probability function for halos of
a given mass and redshift. Within each step a small randomization
assigned the final redshift and mass to the merger fragment. To
calibrate the zero point between the circular velocity (V$_c$, which
the White \& Frenk formulae give) and mass, we normalized to the Milky
Way.  We took V$_c =$ 220 km/s, mass equal to 10$^{11}$ M$_{\odot}$,
and White \& Frenk formulae to obtain M$_{gal =}$ 10$^{11}\,
(\frac{V_c}{220\, km/s})^3$.

The only parameters in the CDM merger models were gas fraction and the
CDM ``bias'' parameter $b$. The ADR statistics from the CDM models
were very high unless the gas fraction was set below 10\%, at which
point the ADev becomes too narrow. The $V$-weighted age and abundance
scatter for the $F_{gas}=0.1$ model is acceptable: 11.4 Gyr and 0.4
dex, respectively.  Increasing the bias parameter helps a little, but
no CDM models were found that matched all statistics simultaneously.
The CDM simulations were also interesting because they predict
variations in Mg-$\sigma$ as a function of galaxy size. For small
galaxies compared to large, CDM predicts similar ADev, younger mean
ages, and much higher asymmetry. Spectroscopic observations
\citep{terle02} indicate a younger mean age for smaller galaxies. Our 
collection of Mg-$\sigma$ observations indicate higher ADev but 
ADR roughly the same.

A variant on the CDM models was to throw out the CDM power spectrum
and adopt a constant power spectrum ($dN \propto M^{-1}$) and return
to a merger probability uniform with time. These models have suitable
ADR and ADev if the gas fraction is set at 2\% - 3\%.  The 2\% model
has a ($V$-weighted) mean age of 11.7 Gyr, with an abundance 
distribution FWHM scatter of .3 dex.  This is the only model
with a mass spectrum (as opposed to the $F_{ini}$ plus constant $dM$
models) that fits all of the observational constraints. Consequently,
even though this constant power spectrum model is not physical, the
good fit seems to indicate that there will be {\em some} power spectrum
that will satisfy the constraints equally well.

Other variants were tried. We tried superimposing two
different schemes atop one another. For instance, suppose half of the
galaxies were truly ancient and the other half was allowed to have
mergers. These superposition schemes fail dramatically. Unless one is
exquisitely artful one ends up with Mg-$\sigma$ diagrams that are
clearly double.

Finally, we tried schemes of variable gas fraction, where the early
universe was assumed to be gas rich, but later mergers are mostly
stellar. Similarly to the regular 4-parameter models, envelope
parameters $n>-0.5$ are ruled out, but uniform-with-time probability
gives good matches for $20 < N_{merger} < 40$. In fact, one such model, 
with F$_{gas}$ varying linearly with time from 0.9 at $z_f$ to 0.1 today 
matches the Mg-$\sigma$ data best of all models. This model is also intriguing 
because it shows strong Mg-$\sigma$ asymmetry evolution at only modest 
redshift (see next section).

\subsection{Predictions for Lookback Studies}

The recent proliferation of 8-12m telescopes will greatly accelerate
the rate at which Mg-$\sigma$ data will become available for distant
galaxies. The result of studies to date, that the drift of the Mg zero
point is ``consistent with passive evolution,'' \citep{Zieg97} will
eventually be broadened to study the shape of Mg residuals as well. We
therefore present some model predictions for high-redshift
studies. There is a major caveat to consider here: model galaxies are
assumed to be ``star forming'' for 50 Myr after a burst, but ``early
type'' at all other times. If merging is a dominant process, then
premerger fragments are often likely to be spiral galaxies. If
semianalytic models like \citet{kauf92} are correct, then, over time,
some ellipticals can accrete a gas disk and become spirals. So the
predictions we show in this section are of questionable validity for
merger-dominated models (i.e. all of those that match the $z=0$
data!), but nevertheless provide a useful first cut for the variety of
behavior we might expect for Mg-$\sigma$ relations at significant
lookback times.

\begin{figure}
\epsscale{0.6}
\plotone{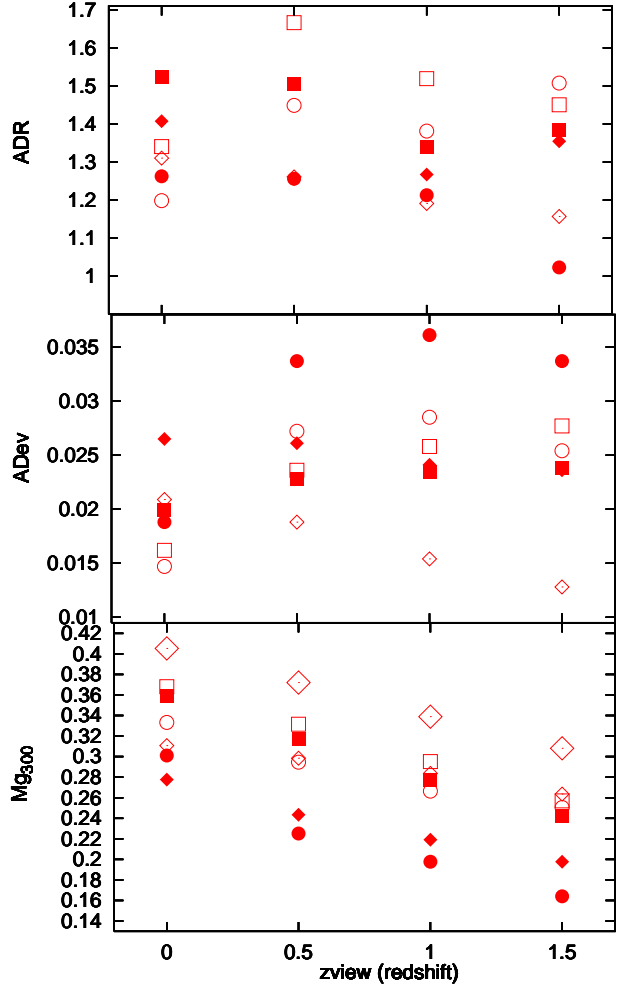}
\figcaption{ Evolution in the past for the ADR (top), ADev (middle)
 and the value of Mg$_2$ at 300 km/s (bottom). Zview is the lookback 
time given in 
redshift. The open squares are for a constant power spectrum model with 
F$_{gas}\, = 2$ \% and the filled squares plot the CDM model with 
F$_{gas}=0.1$.
 The bullets represent the variable gas fraction model with 30 mergers and 
a uniform probability (filled for F$_{ini}=$ 0, open for F$_{ini} =$ 0.5).
 The diamonds plot the regular model with a uniform probability, 40 mergers 
and  F$_{gas}\, =\,0.25$ (filled for F$_{ini} =$ 0, open for 
F$_{ini}=$ 0.5) .
 The bigger open diamonds represent the evolution for F$_{ini}=$ 1. 
\label{fig8}}
\end{figure}

\placefigure{fig8}

Figure \ref{fig8} shows a collection of near-best models as they would
appear if observed at large lookback times. The merging trees of the
model galaxies are computed to redshift zero, but only the partially
complete history of each galaxy is considered for nonzero redshift. So
the further back in time we look, the more incomplete the galaxies
are. Figure \ref{fig8} at a glance shows that Mg-$\sigma$ relations
survive intact even under fairly severe merging scenarios back to when
the Universe was a quarter of its present age. 

The bottom panel shows Mg$_{300}$ as a function of redshift. Within a
zeropoint shift, all models show Mg evolution very similar to the
``ancient, passive evolution'' case shown in large diamond symbols at
strong Mg strength. It is easy to shift any given model track by
adjusting the assumed metallicity scale, so the overall Mg level by
itself is not a strong constraint, but the relative time evolution
should be fairly well modeled. The most shallow slope, that is, the
most constant Mg level with time, is given by the $F_{gas}=0.25$, 
40-merger, uniform probability, $F_{ini}=0.5$ model (small open
diamonds). This is a merger-heavy model. We think that some
researchers will be surprised by the fact that a merger-driven model
is redder, relatively speaking, at $z=1.5$, than pure passive
evolution scenario because one usually thinks of mergers as causing
more star formation activity and hence bluer colors and weaker line
strengths. In reality, it can go either way depending on the
model. The steepest Mg track is from a very similar model but with the
variable $F_{gas}$ option and $F_{ini}=0$ (filled bullets). Neither
steepest nor shallowest models differ by more than 0.05 mag in Mg$_{300}$
from the passive evolution case if they are normalized to match at one
end or the other of the illustrated redshift range.

The tightness and asymmetry of the Mg-$\sigma$ relation also do not
show large changes with lookback time. Both the ADR and ADev can
increase or decrease according to the model, but at a very modest
level that will be hard to measure without a very large sample of
distant galaxies.

\section{Summary and Synthesis}

In this paper we analyze the Mg-$\sigma$ relation in elliptical and S0
galaxies by comparing data and models. 

We adapted an ``ADR'' statistic for the measuring residual asymmetry
and we used the ADev statistic to measure the width of the scatter.
For the combination of all data sets and for 150 $<\sigma <$ 320 km
s$^{-1}$, ADev $=$ .0206 $\pm$ .0014, ADR $=$ 1.265 $\pm$ 0.058 and
Mg$_{300} =$ 0.329 $\pm$ .0039 . The width of the distribution
increases at low $\sigma$. For 50 $<\, \sigma \,<$ 150 km s$^{-1}$ the
ADev $=$ .0282 $\pm$ .0035. The errors of the ADev and ADR statistics
are roughly consistent with the random errors one would expect from
finite number counting as estimated from running simulations multiple
times (the observed ADR scatter is even somewhat lower that what we
would expect).

We also fit the central peak of the Mg residual distribution with a
Gaussian, thinking that this would be a useful number to know under an
ancient formation hypothesis where all of this scatter was imprinted
at formation.  We found a width for the central peak of the Mg
residual $\sigma_g =$ .014 $\pm$ .002 mag, for Mg$_2$ while the
observational error was $\sigma_{obs} = 0.010$. So if total scatter is
a quadrature sum of observational and intrinsic scatter, that is,
$\sigma^2_g = \sigma^2_{intrinsic}\, +\, \sigma^2_{obs}$, we find
$\sigma_{intrinsic} = .01\, mag$. After performing this exercise we
realized that there would be little difference for models that began
with the entire scatter or just the observational portion of the
scatter, and this turned out to be true: the ``ancient formation''
models were not dramatically different from the ``maximum merger'' models.

Merger models were constructed to simulate the Mg-$\sigma$
relation. The merger fragments that join to form the final galaxy were
assumed to be equal in mass. Mergers occur randomly in time (uniform
probability distribution) or randomly under a probability envelope of
the form $(1+z)^n$, where $n$ is a free parameter. The other three
main parameters are the number of mergers, the gas fraction of each
event, and the fraction of the galaxy that formed in the distant past,
at $z=z_f=5$ for this paper. Also included was an artificial scatter
in Mg strength. This scatter was set equal to the observational error
for the ``maximum merger'' hypothesis, and equal to the entire
symmetric scatter for the ``ancient formation'' hypothesis, but since
the latter was only 40\% larger than the former, the two hypotheses
generated similar results.

The models predict a simulated Mg-$\sigma$ diagram, from which we
compute Mg$_{300}$, ADev, and ADR in the same way as we did for the
observational data. We also output V-weighted mean age and an
abundance distribution since observational constraints also exist on
these quantities. Some immediate conclusions are:

$\bullet$ A small number (5-20) of (severe) mergers leads to scatter
and asymmetry larger than observed.

$\bullet$ To the contrary, with many mergers (60-100) a large number
of small events leads to small scatter and small asymmetry because
merging histories are similar. Most of these models showed {\em too}
tight of a relation, and the gas-rich ones also tended toward
unacceptably young mean ages.

$\bullet$ For the merger probability envelope, a positive $n$ in
$(1+z)^n$ concentrates galaxy formation to the early universe, making
the Mg-$\sigma$ relation thin.  Because we find no successful models
for $n>0$, we clash with many published merger rate results such as
$(1+z)^{2.3\pm0.2}$, $(1+z)^{2.3\pm 0.9}$, $(1+z)^{4.5}$, or
$(1+z)^{3.1{\rm \ or \ } 2.5}$ \citep{conseli01,patt02,okam00,gove99}
but we agree with a few such as $(1+z)^{0.1\pm 0.5}$
\citep{carl00}. Our $n<0$ result is very robust, so we presume that
the difference between this result and others stems from differences
in treatment (i.e. redshift range, galaxy types, definition of merger
rate, observables). For example, \citet{kauf96} find a $(1+z)^{1.5}$
dependence for early type galaxies, but this is for number evolution,
a quantity that we do not model because it requires tranformations to
other Hubble types.

$\bullet$ An increased gas fraction increases the number of new
stars formed in each merger. This increases the scatter in Mg strength
but increases asymmetry only on some models.
 
$\bullet$ If a larger fraction of the galaxies form at $z=z_f$, both 
scatter and asymmetry decrease.

Finally, let us briefly discuss the meaning of these results.  Our
most successful models involved a moderate number of mergers (20-60)
with mergers equally probable over cosmic time. The model with a
constant gas fraction that fit the best was with $F_{gas}=0.15$-0.3 but a
model with gas fraction that varies linearly with time from 0.9 at
$z_f$ to 0.1 at the present epoch fit even better. One other model, a
variable-mass scenario with a flat power spectrum and merger
probability constant with time fit the constraints if the gas fraction
was 2\% - 3\%. The overall yield of successful models was far below
our initial expectations; we thought many solutions would be
possible. This is good news for understanding galaxy evolution: tight
constraints are better than loose ones.

We also find some clarity for the basic question, ``Are ellipticals
old or young?'' To arrive at age information, one must look at the
most sensitive indicators. A brief perusal of \citet{wor94} or other
population model paper weighted by the inverse of the observational
error orders the list, from most sensitive to least sensitive:
Balmer-metal diagrams, Mg-$\sigma$ relation, Fundamental Plane, and
color-magnitude diagram. Thus Balmer-metal diagrams show very large
age scatter that is harder to detect with other methods. The present
study of the Mg-$\sigma$ relation finds that the models that fit every
observable have an age range of 7-10 Gyr. This does not force a
revision of the Balmer-based age scale, but it keeps alive the
long-held suspicion that the Balmer ages are all somewhat too
young. The Balmer-derived age scatter is fully supported by our study.

(1) Balmer-metal and Mg-$\sigma$ diagrams are compatible, both
    implying considerable age scatter and a mean age (7-10 Gyr) within
    the uncertainty of the Balmer method.  The tightness of the
    Mg-$\sigma$ relation is compatible with the less-sensitive FP and
    color-magnitude diagram tightness, and both are compatible with
    the Balmer-metal diagrams. In short, we see no barriers to fairly
    strong merger hypotheses.

(2) In contrast, we find little support for the ancient merger hypothesis:
    the tamest merger scenarios that we could find are still much
    wilder than quiescent evolution since $z_f$. The least active model
    (2-3\% gas fraction and flat mass spectrum) still involves massive
    mergers (albeit without much gas) up to the present time. On the
    other hand, this model's mean age was 11 Gyr and it was still able
    to match the width and asymmetry of the Mg-$\sigma$ relation, so
    there may be a related model that can satisfy a relaxed version of
    the ancient merger hypothesis.

We found this avenue of investigation unexpectedly fruitful in the
sense that very strong observational constraints allow only very
select interpretations to survive. Future theoretical work should
involve more choices for mass power spectrum as a first priority,
although many aspects of the modeling could be improved. Observational
work at intermediate redshift is also needed, although our results
regarding high quality and large $N$ of such studies make this job
more challenging.

\begin{acknowledgements}

We thank M. Bernardi, D. Burstein, M. J. Hudson, and P. Prugniel for
machine-readable Mg-$\sigma$ data. Washington State University
provided partial funding for this effort.

\end{acknowledgements}

\clearpage

\begin{deluxetable}{lrrrrrrrr} 
\tabletypesize{\footnotesize}
\tablecolumns{9} 
\tablewidth{0pc} 
\tablecaption{Statistics on the Different Data Sets\label{tab-1}} 
\tablehead{ 
\colhead{Paper}    & \colhead{N}       & \colhead{N$^{'}$} & 
\colhead{Mg 300}   & \colhead{Std. Dev.} & 
\colhead{Skewness} & \colhead{ADev} &
\colhead{ADR} & \colhead{$\sigma_{g}$ }
} 
\startdata
\multicolumn{9}{c}{\textbf{Mg$_2$}}\\  
\cline{2-9}
Prugniel, Simien 1996& 644 & 246& 0.314 & 0.023 & 
 -1.017 & 0.017 & 1.120 & 0.013\\ 
Trager et al. 1998 (Lick)& 256& 183 & 0.339 & 0.032 & 
 -1.750 & 0.022 & 1.544 & 0.014\\ 
Efar 1999&140 &130 &0.331 &0.021 & 
-0.324 & 0.017 & 1.210 & 0.013 \\ 
Dressler et al. 1991&123 &107 &0.318 &0.023 & 
-0.722 &0.018 &1.538 &0.011\\ 
Bernardi et al. 1998 (all)& 721&654 &0.329  & 0.033 & 
 -0.184 &0.025 & 1.230 &0.020\\ 
Bernardi et al. 1998 (field)& 469&366 &0.331  & 0.033 & 
 -0.186 & 0.025& 1.151 &0.020\\ 
Bernardi et al. 1998 (group)& 104&91 & 0.333 & 0.031 & 
 -0.369 & 0.024& 1.361& 0.015\\ 
Bernardi et al. 1998 (cluster)& 104&83 & 0.331 & 0.032 & 
 -0.052 & 0.023& 1.128&0.013\\
Hudson 2001 &528 &391&0.332& 0.018&
-0.369 & 0.014& 1.105& 0.013\\
\multicolumn{9}{c}{\bf{Mg$b$}} \\
\cline{2-9}
Trager et al. 1998 (Lick) & 256 & 183 & 5.091 & 0.526 & 
 -0.971 & 0.378 & 1.220 &0.2\\ 
Efar 1999 &164 &144 &5.260 &0.422 & 
-0.573 & 0.314 & 0.990& 0.3 \\ 
\multicolumn{9}{c}{\bf{$<$Mg$_2$$>$\tablenotemark{a}}}  \\
\cline{2-9}
Trager et al. 1998 (Lick)& 256&183 & 0.338& 0.032 & 
 -2.472 & 0.021 & 1.788 &0.011\\ 
7-Samurai & 577&358 & 0.324& 0.038 & 
 -5.545 & 0.020 & 1.609 &0.015\\ 
\enddata
\tablecomments{Col. (1)---Reference. 
Col. (2)---Number of galaxies in the sample. 
Col. (3)---Number of galaxies used for the statistic. 
Col. (4)---Value of the Mg index at $\sigma = 300\, km/s$. 
Col. (5)---Standard deviation. 
Col. (6)---Skewness. 
Col. (7)---ADev or Average Deviation. 
Col. (8)---ADR or Average Deviation Ratio. 
Col. (9)--- $\sigma$ from the (Gaussian) fit.}
\tablenotetext{a}{$<Mg_2>=$0.6\,Mg$_2$ $+$ 0.4\,Mg${^{'}}_2$, with 
Mg${^{'}}_2$ $=$ 0.03 $+$ 2.10\,Mg$_1$ $-$62\,Mg$^4_1$}
\end{deluxetable} 

\clearpage
\clearpage
\clearpage

\clearpage

\clearpage

\clearpage

\clearpage

\clearpage

\clearpage

\clearpage

\clearpage

\clearpage

\clearpage

\clearpage

\clearpage

\end{document}